\documentclass[
  journal=jpccck,
  manuscript=article,
  layout=twocolumn]{achemso}

\usepackage[fontsize=10pt]{scrextend}
\usepackage[T1]{fontenc}
\usepackage[utf8]{inputenc}
\usepackage{filecontents}
\usepackage{xr}
\usepackage[separate-uncertainty=true]{siunitx}
\usepackage{amsmath}
\usepackage{mathtools}
\usepackage[colorlinks=true,
    citecolor=blue,
    linkcolor=blue,
    urlcolor=blue,pagebackref=false]{hyperref}
\usepackage[version=3]{mhchem}
\usepackage[pagewise]{lineno}
\usepackage{fourier}

\setkeys{acs}{doi = false}

\graphicspath{
  {Figs/}  
  }

\newcommand{\st}[1]{_{\mathrm{#1}}} 
\SectionNumbersOn
\DeclareSIUnit{\molar}{M}
\DeclareSIUnit{\sq}{sq.}
\renewcommand{\Re}{\operatorname{Re}}
\renewcommand{\Im}{\operatorname{Im}}

				\title{Light-dependent Impedance Spectra and Transient Photoconductivity in a Ruddlesden--Popper 2D Lead-halide Perovskite Revealed by Electrical Scanned Probe Microscopy and Accompanying Theory}
        \author{Ali Moeed Tirmzi}
        \affiliation{Department of Chemistry and Chemical Biology, Cornell University, Ithaca, New York 14853, United States}
        \author{Ryan P. Dwyer}
        \affiliation{Department of Chemistry and Biochemistry, University of Mount Union, Alliance, Ohio 44601, United States}
        \author{Fangyuan Jiang}
        \affiliation{Department of Chemistry, University of Washington, Seattle, Washington 98195, United States}
        \author{John A. Marohn}
        \affiliation{Department of Chemistry and Biochemistry, University of Mount Union, Alliance, Ohio 44601, United States}
        \email{jam99@cornell.edu}

\makeatletter\@input{file_2.tex}\makeatother
\begin{document}

\sloppy 

\begin{abstract}
Electric force microscopy was used to record the light-dependent impedance spectrum and the probe transient photoconductivity of a film of butylammonium lead iodide, \ce{BA2PbI4}, a 2D Ruddlesden--Popper perovskite semiconductor.
The impedance spectrum of \ce{BA2PbI4} showed modest changes as the illumination intensity was varied up to \SI{1400}{\milli\watt\per\square\centi\meter}, in contrast with the comparatively dramatic changes seen for 3D lead-halide perovskites under similar conditions.
\ce{BA2PbI4}'s light-induced conductivity had a rise time and decay time of $\sim$\SI{100}{\micro\second}, \SI{e4}{} slower than expected from direct electron-hole recombination and yet \SI{e5}{} faster than the conductivity-recovery times recently observed in 3D lead-halide perovskites and attributed to the relaxation of photogenerated vacancies.
What sample properties are probed by electric force microscope measurements remains an open question.
A Lagrangian-mechanics treatment of the electric force microscope experiment was recently introduced by Dwyer, Harrell, and Marohn which enabled the calculation of steady-state electric force microscope signals in terms of a complex sample impedance.
Here this impedance treatment of the tip-sample interaction is extended, through the introduction of a time-dependent transfer function, to include time-resolved electrical scanned probe measurements.
It is shown that the signal in a phase-kick electric force microscope experiment, and therefore also the signal in a time-resolved electrostatic force microscope experiment, can be written explicitly in terms of the sample's time-dependent resistance (\emph{i.e.}, conductivity).
\end{abstract}


\section{Introduction}
Recent interest in using 2D Ruddlesden--Popper perovskite for photovoltaic applications has increased due to unique properties that distinguish them from the widely studied 3D perovskite.
These properties include higher environmental stability due to the hydrophobic nature of the organic spacer and higher formation energy \cite{Quan2016feb}, increased exciton binding energy due to alternating organic and inorganic layer with disparate dielectric constants \cite{Straus2018feb}, surprisingly reduced in-plane ion motion both in the dark and under illumination presumably due to higher formation energy of vacancies \cite{Xiao2018feb}, and the existence of edge states with low energy that are thought to help in exciton dissociation or lead to unusual charge carrier densities \cite{Blancon2017mar,Wang2019jul}.
DC galvanostatic polarization measurements and AC impedance spectroscopy have revealed 
\ce{(C6H10N2)PbI4}, a commonly studied 2D perovskite, is a mixed ionic electronic conductors with low dark ionic and electronic conductivities on the order of \SI{1e-10}{\siemens\per\centi\meter} \cite{Lermer2018aug}. 
Despite the presence of the insulating organic layer and exciton binding energy of $100$'s of milli electron volts, free carrier generation in the quasi 2D perovskites with more than one layer of the inorganic octahedron (\textit{n} $>$ $1$ where \textit{n} is the number of layers of the inorganic octahedron) seems to be efficient enough such that power conversion efficiency greater than $18 \%$ have been reported for solar cells made from a quasi-2D perovskite \cite{Yang2018oct}.
The working principle of such cells for higher \textit{n} is thought to involve charge transfer from layered 2D regions to a 3D perovskite network \cite{Lin2019mar}.
An alternative proposal for the efficient generation of free carriers from bound exciton involves edge states \cite{Blancon2017mar}.
Anisotropy between in-plane and out-of-plane charge carrier mobility has been reported; the charge carriers motion is not inhibited completely in the out-of-plane direction, with reported mobility on the order of $\SI{1e-4}{\centi\meter\squared\per\volt\per\second}$ \cite{Lin2019mar} for an $n = 3$ system.
Here we study a 2D Ruddlesden--Popper phase of perovskite butylammonium lead iodide (\ce{(C4H9NH3)2PbI4} or \ce{BA2PbI4}, herein referred BAPI).
BAPI, with $n = 1$, where higher dimension/bulk perovskite is absent, can act as a model system to understand the true nature of charge motion in a quasi 2D system.

We have recently revealed persistent photoinduced conductivity in a variety of 3D perovskite samples using time- and frequency-resolved electrical SPM techniques \cite{Tirmzi2017jan, Dwyer2019mar, Tirmzi2019jan}.
The conductivity dynamics in these 3D systems, while substrate dependent, generally exhibited a slow, activated recovery in the dark around room temperature which became faster at low temperature.
We hypothesized that light-induced vacancy generation could explain the long lived changes in conductivity \cite{Tirmzi2017jan, Tirmzi2019jan}.
We wanted to test if the same surprisingly persistent light-induced conductivity was present in the 2D system.

Time resolved electrical scanning probe microscopy (SPM) techniques have been used to probe charge dynamics in photovoltaic and ion conducting materials \cite{Mascaro2019mar}.
These techniques combine the very high charge sensitivity of microcantilevers  with carefully designed measurement protocols to push the time resolution of measured sample parameters (such as capacitance, non-contact friction, dissipation, and surface (photo) voltage) beyond the natural time resolution limit of the cantilever period. 
To harness the true power of these microscopic and localized measurements, and compare results with bulk electrical characterization techniques, it is imperative that we establish which sample properties these measurements probe \cite{Harrison2018dec}.
By extending a recently developed theoretical framework \cite{Dwyer2019jun} for quantitatively interpreting electrical scanning probe microscopy data, we show  here that changes in sample resistance can be followed by combination of techniques that measure sample response either in the frequency and/or in the time domain. 

Ginger and coworkers have recently performed one of the first electrical SPM studies of the 2D Ruddlesden--Popper phase of perovskite BAPI to characterize light-induced charge carrier dynamics. 
They used two electrical scanning probe techniques, namely fast-free time-resolved electrostatic force microscopy (FF-tr-EFM) and general-acquisition-mode Kelvin probe force microscopy (G-mode KPFM, or G-KPFM).
They measured charging times and time-resolved surface photovoltage measurements, observing $100$ to \SI{1000}{\micro\second} dynamics in these material in response to light irradiation over a large area \cite{Giridharagopal2019feb}.
They attributed these dynamics to trap-mediated electron or ion motion. 
The two techniques measured time scales that were different by several hundred microseconds.
Dielectric response in 3D perovskites is known to be highly frequency dependent where the low frequency response ($\leq$ $\SI{10}{\kilo\hertz}$) is dominated by ion motion \cite{Wilson2019jan}.
Therefore, we expect that measurements that sample different parts of the frequency response may have a different temporal dependence in response to an external stimuli. 

Here we use time-resolved EFM (tr-EFM) \cite{Coffey2006sep}, phase-kick EFM (pk-EFM) \cite{Dwyer2017jun}, broadband local dielectric spectroscopy (BLDS) \cite{Labardi2016may,Tirmzi2017jan}, and dissipation microscopy \cite{Cox2016,Tirmzi2017jan,Tirmzi2019jan} to probe changes in the sample in response to optical irradiation.
We show that the transfer function representation of SPM used to describe broadband local dielectric spectroscopy and dissipation measurements can be extended to understand the origin of the temporal response measured in time resolved tr-EFM and pk-EFM measurements \cite{Dwyer2019jun}.
We show that these measurements probe changes in sample resistance instead of the commonly assumed capacitance changes in 2D BAPI.
We find that light-induced conductivity in the 2D BAPI system recovers promptly, in marked contrast to the slow, activated recovery observed in the many 3D perovskite systems studied to date.


\section{Experimental Section}

\subsection{Scanning probe microscopy}

All experiments were performed under vacuum (\SI{5e-6}{\milli\bar}) in a custom-built scanning Kelvin probe microscope described in detail elsewhere \cite{Dwyer2017jun}.
The cantilever used was a MikroMasch HQ:NSC18/Pt conductive probe.
The resonance frequency and quality factor were obtained from ringdown measurements and found to be $\omega\st{c}/2\pi = f\st{c} = \SI{60.490}{\kilo\Hz}$ and $Q = \num{28000}$ respectively at room temperature.
The manufacturer's specified resonance frequency and spring constant were $f\st{c} = 60$ to $\SI{75}{\kilo\Hz}$ and $k = \SI{3.5}{\N\per\m}$.
Cantilever motion was detected using a fiber interferometer operating at $\SI{1490}{\nano\meter}$ (Corning model SMF-28 fiber).
The sample was illuminated from above with a variable-intensity $\SI{405}{\nano\meter}$ diode laser.
More experimental details regarding the implementation of broadband local dielectric spectroscopy, tr-EFM, and pk-EFM can be found in the Supporting Information.

\subsection{Sample preparation}
Thin film of \ce{BA2PbI4} was synthesized on ITO adapted from the procedure reported in Ref.~\citenum{Giridharagopal2019feb}.
The film was approximately \SI{500}{\nano\meter} thick.
Based on the absorption coefficient, we expect the absorption length to be $\approx$ $\SI{150}{\nano\meter}$ which is much shorter than the film thickness. 
Briefly, precursor solution was prepared by dissolving $\SI{1.8}{\molar}$ butylammonium iodide (BAI) and $\SI{0.9}{\molar}$ lead (II) iodide (\ce{PbI2}) in $\SI{1}{\milli\liter}$ of dimethylformamide (DMF). 
Perovskite solution was spin coated on the ITO substrates at 4000 rpm for $\SI{40}{\second}$ (with high acceleration), followed by thermal annealing for $\SI{10}{\minute}$ at $\SI{100}{\celsius}$. 
A representative AFM, absorption spectra, and XRD spectra is included in the supporting information.


\section{Results}

\subsection{Theoretical background}
\label{Sec:theoretical-background}
In this section, we extend our previously developed impedance model for SPM experiments\cite{Tirmzi2017jan,Dwyer2019jun} to the situation where the sample properties are time-dependent.
Our new model allows us to interpret all of our SPM data in a common framework.
We explain the key features of this model below; a detailed derivation is included in the Supporting Information (\ref{sec:Theory-derivation}).

We investigate the sample's time-dependent properties in the dark and under illumination using SPM experiments.
To interpret these experiments, we need to connect the observed cantilever amplitude, frequency, and phase back to the sample properties.
These observable depend on the tip-sample force $F\st{ts} = C\st{tip}'(x) q\st{t}^2/(2C\st{tip}(x)^2)$ where $C\st{tip}$ is the tip capacitance, $C\st{tip}' = d C\st{tip}/dx$ is the derivative of the tip capacitance with respect to the vertical direction, $x$ is the cantilever displacement, and $q\st{t}$ is the tip charge. 
(In keeping with the conventional notation for one-dimensional harmonic oscillators and to be consistent with the notation used in Ref.~\citenum{Dwyer2019jun}, we represent cantilever displacement with $x$ rather than $z$, with increasing $x$ corresponding to motion of the cantilever tip away from the sample surface.)
The component of the tip-sample force evolving in phase with tip motion shifts the cantilever frequency, $\Delta f\st{c}$, while the out-of-phase force component causes dissipation, $\Gamma\st{s}$, which in turn changes the cantilever amplitude.
Tip charge $q\st{t}$ depends on tip capacitance and sample impedance (Eq.~\ref{eq:first-order-tip-charge} and Eq.~\ref{eq:first-order-tip-charge-time}).
In Refs.~\citenum{Tirmzi2017jan} and \citenum{Dwyer2019jun}, we used an impedance model and perturbation theory to develop an accurate steady-state approximation for $q\st{t}$ and the cantilever observables.
In this model, the central quantity that links sample properties to cantilever observables is the transfer function $H(\omega) = V\st{t} \big/ V\st{ts}$.
Here $V\st{t}$ is the voltage dropped across the tip-sample gap; $V\st{ts}$ is the applied external voltage, dropped across both the tip-sample gap and the sample; and $\omega$ is a modulation frequency.
The tip capacitance $C\st{tip}$ and sample impedance $Z(\omega)$ together determine the transfer function, $H(\omega) = (1 + j \omega C\st{tip} Z(\omega))^{-1}$.
The central results of this model are that $\Delta f \propto \Re{\! \big[H(\omega_0)\big]} (V\st{ts} - \phi)^2$ and $\Gamma\st{s} \propto \Im{\! \big[H(\omega_0)\big]} (V\st{ts} - \phi)^2$ with $\omega_0 = 2 \pi f_c$ the cantilever frequency and $\phi$ the surface potential (Eqs.~7 and 8 in Ref.~\citenum{Dwyer2019jun}).
The signal $\alpha$ in the broadband dielectric spectroscopy (BLDS) measurement described below can likewise be expressed in terms of the transfer function: $\alpha \propto |H(\omega\st{m})|^2 (V\st{ts} - \phi)^2$, with $\omega\st{m}$ the voltage-modulation frequency (Eq.~10 in Ref.~\citenum{Dwyer2019jun}).
These expressions straightforwardly connect sample resistance and capacitance, through $Z$ and $H$, to changes in cantilever frequency, dissipation, and BLDS signal.

The above description is only appropriate when the sample properties are time-independent and the experiment is conducted at steady-state.
Yet we are also interested in analyzing time-dependent changes in sample properties in response to illumination.
For this reason, we extend the theoretical treatment to the case where the sample parameters may be time-dependent.
Assuming the applied voltage or light does not induce oscillations of the tip charge at the cantilever frequency, our results for time-independent systems can be extended to time-varying systems by replacing the time-independent transfer function $H(\omega)$ with the time-varying transfer function $H(\omega, t)$ defined below \cite{Kailath1980book}.
Using these assumptions, we derive a general result for the frequency shift in an electric or Kelvin probe probe microscopy experiment
\begin{equation}
\Delta f(t) = -\frac{f\st{c} }{4 k\st{c}} \left ( C''_q + \Delta C'' \Re{\! \big[H(\omega_c, t)\big]}\right ) (V\st{t}(t) - \phi)^2
\label{eq:frequency_shift_time_varying}
\end{equation}
where $f\st{c}$ is the cantilever resonance frequency, $k\st{c}$ is the cantilever spring constant, $C''_q$ and $\Delta C''$ are related to the second derivative of the tip capacitance, $\phi$ is the surface potential that represents internal voltage source, and $V\st{t}(t)$ is the tip voltage drop calculated under the assumption that the tip displacement is fixed at zero.
Since $\phi$ $<$ $V\st{t}$ (Fig.~\ref{fig:surface-potential}), we ignore $\phi$ to simplify our analysis.
This result allows pk-EFM and tr-EFM, experiments that inherently involve time-dependent dynamics, to be analyzed using the same framework as the steady state measurements considered previously with this theory. 
The applicability of this approximation to the experiments described here is illustrated by the close agreement between numerical simulations of the cantilever frequency and phase and the analytical approximation developed here (Fig.~\ref{fig:simulated-EFM} and Fig.~\ref{fig:compare-phase-shift}).

\subsubsection{Transfer function}

In this paper, we model the sample using a parallel resistor $R\st{s}$ and capacitor $C\st{s}$. If $R\st{s}$ and $C\st{s}$ are time-independent, the transfer function $H$ is
\begin{equation}
H(\omega) 
	= \frac{1+ j\omega R\st{s} C\st{s}\hphantom{+C}}{1 + j \omega R\st{s} (C\st{s} + C\st{tip})} = \frac{1 + j g^{-1}\omega \omega\st{fast}^{-1}}{1 + j\omega \omega\st{fast}^{-1}}
\label{eq:H-full}
\end{equation}
where $\omega\st{fast}(t)= (R\st{s}C\st{tot})^{-1}$ is the time-independent frequency and $g = C\st{tot} / C\st{s}$ where $C\st{tot} = C\st{tip} + C\st{s}$ is the total capacitance.
$C\st{tip}$ is the capacitance between tip and a hypothetical ground plane located at the sample surface and $C\st{s}$ is the sample capacitance when the sample impedance is modeled as a parallel resistor and a capacitor.
The complex-valued transfer function in Eq.~\ref{eq:H-full} has a real part which determines the in-phase forces and an imaginary part which determines the out-of-phase forces acting on the cantilever. 
An increase in cantilever dissipation is caused by out-of-phase forces acting on the cantilever arising from changes in the value of the imaginary part of the transfer function  at the cantilever frequency $\omega\st{c}$ \cite{Tirmzi2017jan}.

The situation is considerably more complicated if $R\st{s}$ and/or $C\st{s}$ vary with time, as they do when the light intensity is changed.
In this case, the time-varying response function is
\begin{multline}
H(\omega, t) = \frac{C\st{tip}}{C\st{tot}} \int_0^{\infty} \exp \left [  - \int_{t-\tau}^{t} \omega\st{fast}(t') dt' \right ] \times
\\
 \omega\st{fast}(t-\tau) e^{-j\omega \tau} \, d \tau
 + \frac{C\st{s}}{C\st{tot}},  
\label{eq:time-varying-response}
\end{multline}
where $C\st{tot} = C\st{tip} + C\st{s}$ is the total capacitance, and $\omega\st{fast}(t)= (R\st{s}C\st{tot})^{-1}$ is the time-dependent frequency at which charge responds.

\subsection{Dissipation measurements}
\begin{figure*}[t]
\includegraphics[width=6.5in]{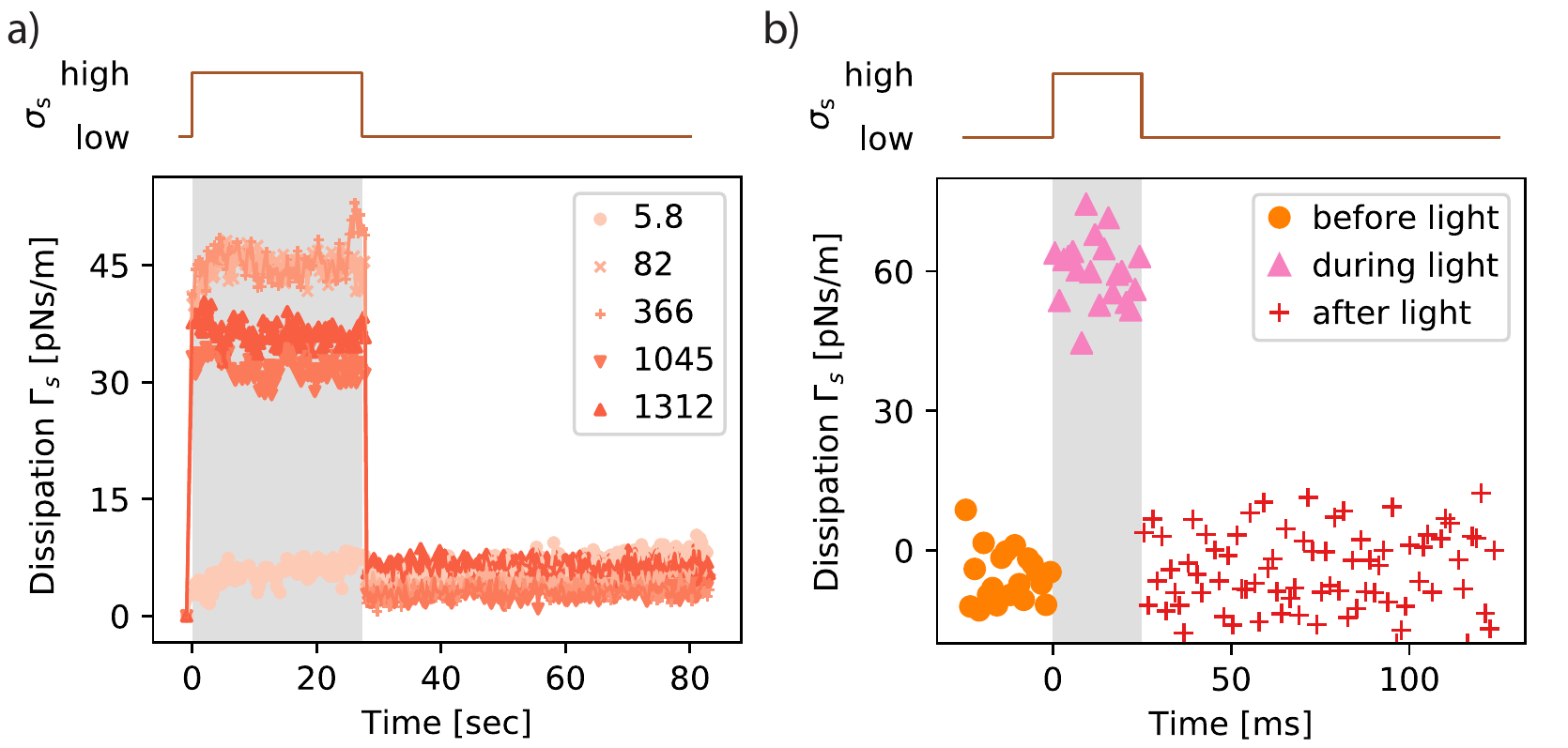}
\caption[Ring-down dissipation measurement on BAPI]{a) Bottom: Sample-induced cantilever dissipation over a BAPI film increases promptly when light is applied (shaded region) and recovers in less than $\SI{100}{\milli\second}$ when light is removed.
Top: The corresponding sample conductivity.
The dissipation is quantified through ring-down measurements at indicated light intensities (in $\SI{}{\milli\W\per\cm\squared}$).
Experimental parameters: $V\st{ts} = \SI{-4}{\V}$, $h = \SI{125}{\nano\meter}$, $\lambda = \SI{405}{\nano\meter}$. 
b) Light-induced sample dissipation (\emph{i.e.} sample conductivity) recovers in less than $\SI{2}{\milli\second}$. 
Data obtained through an implementation of tr-EFM with a constant DC bias applied, as described in the text. 
Experimental parameters: $I_{h\nu} = \SI{120}{\milli\W\per\cm\squared}$,  $V\st{ts} = \SI{-6}{\V}$, $h = \SI{200}{\nano\meter}$, $\lambda = \SI{405}{\nano\meter}$.}
\label{fig:dissipation-ringdown}
\end{figure*}

We start our investigation by measuring light induced cantilever dissipation.
Dissipation tracks changes in the sample resistance $R\st{s}$ and sample capacitance $C\st{s}$ when the associated time constant $R\st{s}(C\st{s} + C\st{tip})$ is near the cantilever period of $\omega\st{c}^{-1}$ which is the case here.
In our previous report, we showed that changes in dissipation caused by light can take tens to hundreds of seconds to recover in 3D perovskite samples \cite{Tirmzi2017jan,Tirmzi2019jan}.
We perform this measurement by recording changes in cantilever ring-down time \cite{Tirmzi2017jan,Tirmzi2019jan}.
A DC voltage of \SI{-4}{\volt} is applied before the start of the measurement to ensure that any slow process in response to the tip electric field in the dark will not interfere with the measurement.
Here we see an increased dissipation when the light is turned on that goes away within the time resolution of the measurement, $\leq \SI{100}{\milli\second}$, at least three orders of magnitude faster than the recovery time seen in 3D \ce{CsPbBr3} and 3D FAMACs films.

To better resolve the associated time constant, we turn to an implementation of the tr-EFM measurement.
The cantilever is allowed to freely ring down and the cantilever oscillation is demodulated to obtain the instantaneous cantilever amplitude, phase, and frequency.
The measurement, shown in Fig.~\ref{fig:dissipation-ringdown}b, proceeds as follows.
A DC voltage of $\SI{-6}{\volt}$ is applied throughout the measurement.
This voltage ensures that any slow tip charge equilibration on application of the tip electric field does not affect the measurement.
The cantilever is allowed to ring down for $\SI{125}{\milli\second}$ starting at time $t = \SI{-25}{\milli\second}$.
At time $t = \SI{0}{\second}$, a $\SI{25}{\milli\second}$ light pulse is initiated.
This process is repeated for $600$ acquisitions and the averaged cantilever amplitude data is divided into $\SI{2}{\milli\second}$ bins.
The known cantilever frequency and initial amplitude is used to extract cantilever $Q$ for each $\SI{2}{\milli\second}$ bin giving a time resolution of $\SI{2}{\milli\second}$. 
This time resolution is $\sim \! 50$ times better than the time resolution in Fig.~\ref{fig:dissipation-ringdown}a.
It is clear from these measurements that the changes in the transfer function $H(\omega\st{c})$ at the cantilever frequency happen in less than $\SI{2}{\milli\second}$.
Analysis performed with $\SI{1}{\milli\second}$ bins gave a similar result, albeit with worse SNR, confirming the sub-millisecond time scale of the underlying charge dynamics that cause dissipation.

\subsection{Broadband local dielectric spectroscopy}
\begin{figure*}[t]
\includegraphics[width=6.5in]{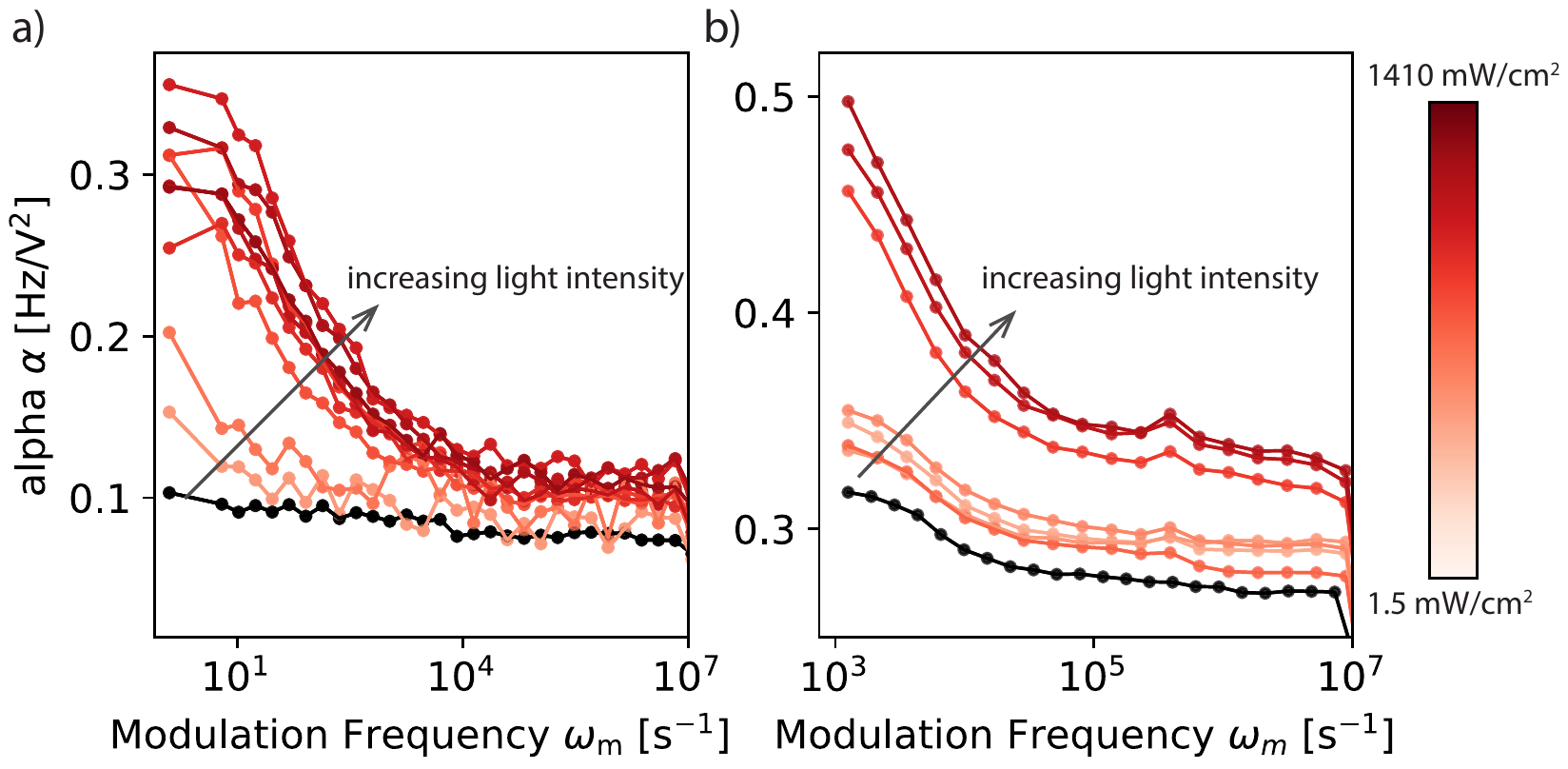}
\caption[Frequency shift BLDS]{a) Frequency-shift broadband dielectric spectra reveals a large increase in the sample's low frequency ($< \SI{1000}{\per\second}$) response when the light intensity is increased.
Experimental parameters: $V\st{pp} = \SI{6}{\V}$, $h = \SI{125}{\nano\meter}$. 
b) High frequency response at indicated illumination intensities resolved by amplitude modulation broadband local dielectric spectroscopy. 
Experimental parameters: $V\st{pp} = \SI{6}{\V}$, $h = \SI{125}{\nano\meter}$.}
\label{fig:frequency-BLDS}
\end{figure*}
Broadband local dielectric spectroscopy (BLDS) can be used to monitor changes in the  impedance of a semiconducting sample over a wide frequency range \cite{Tirmzi2017jan, Tirmzi2019jan}.
In a BLDS spectrum, we plot a voltage normalized frequency shift alpha ($\alpha$) with units of $\SI{}{\hertz\per\volt\squared}$ versus the modulation frequency of the applied tip-sample voltage ($2 \pi \, f\st{m}$).
We turn to two different implementations of broadband local dielectric spectroscopy, \textit{i.e.}\ 1) average frequency-shift BLDS and 2) amplitude-modulation BLDS, to map out the transfer function over a wide frequency range.
In the first implementation, described in detail elsewhere \cite{Tirmzi2017jan}, we record the average frequency shift as a function of applied voltage modulation frequency, with a 1:1 ``modulation on'' to ``modulation off'' duty cycle employed to allow lock-in detection.
While the SNR of this measurement is relatively low compared to the amplitude-modulation BLDS shown in the proceeding figure, this measurement allows us to access the low frequency part of the transfer function below \SI{1000}{\per\second}.
In Fig.~\ref{fig:frequency-BLDS}, we can see that the transfer function changes in the low frequency region are much more obvious than the small changes in the near-cantilever-frequency region.
This observation is in line with the relatively small dissipation changes measured in Fig.~\ref{fig:dissipation-ringdown}b.
We note that in the dark, the roll off of the transfer function is below the lowest probed frequency here (\SI{1}{\per\second}).

To better resolve changes in the high frequency region, we turn to amplitude-modulated BLDS, Fig.~\ref{fig:frequency-BLDS}b.
Different numerical values of $\alpha$ in Fig.~\ref{fig:frequency-BLDS}a and Fig.~\ref{fig:frequency-BLDS}b are attributed to different regions in the sample.
This measurement's lock-in detection scheme allows us to clearly resolve changes in the dielectric spectra in the frequency range of kHz to MHz that were not obvious in Fig.~\ref{fig:frequency-BLDS}a.
In Fig.~\ref{fig:frequency-BLDS}b, we can see that increasing light intensity increases the value of $\alpha$. 
To describe changes in BLDS spectra, we have previously used an $RC$ electrical model for the sample with one resistor and one capacitor. 
While a simple $RC$ is often not enough to capture the full impedance changes in a mixed ionic-electronic conductor \cite{Tirmzi2019jan}, it is nevertheless a useful starting point to understand physical process causing the changes in the impedance of the sample.
In our model, the measured $\alpha$ is proportional to $H(\omega)$ which is a function of two parameters  
\begin{equation}
\omega\st{fast}^{-1}  = R\st{s} (C\st{s} + C\st{tip}) \hspace{0.5em} \text{and} \hspace{0.5em} 
g = C\st{tot} / C\st{s}.
\label{eq:tau-g}
\end{equation}
with $\omega\st{fast}^{-1}$ a time constant describing transfer function roll-off.
From Eq.~\ref{eq:H-full}, the high-frequency limit of $H(\omega)$ is $1/g$, Eq.~\ref{eq:tau-g}.
The high-frequency data in Fig.~\ref{fig:frequency-BLDS} give $g \sim 2$ to $3$, indicating that the tip capacitance and sample capacitance are of similar size.

Light-induced changes in the BLDS spectra can be attributed to resistance and/or capacitance.
The BLDS spectra show $g$ decreasing slightly with increasing light intensity.
Taking derivatives, $\delta g = - C\st{tip} C\st{s}^{-2} \, \delta C\st{s}$.
If $C\st{tip}$ is independent of light intensity, the observed decrease in $g$ thus indicates a slight increase in $C\st{s}$ with increasing light intensity.
The BLDS spectra also show a time constant $\omega\st{fast}^{-1}$ decreasing from seconds to milliseconds as light intensity increases from \SI{1.5}{} to \SI{1410}{\milli\watt\per\square\centi\meter} (and pinned at a near constant value at high light intensity).
This decrease in time constant requires a decrease in the $R\st{s} (C\st{s} + C\st{tip})$ product, which can then only be explained by a large decrease in $R\st{s}$ with increasing light intensity.

\begin{figure}[t]
\includegraphics[width=3.25in]{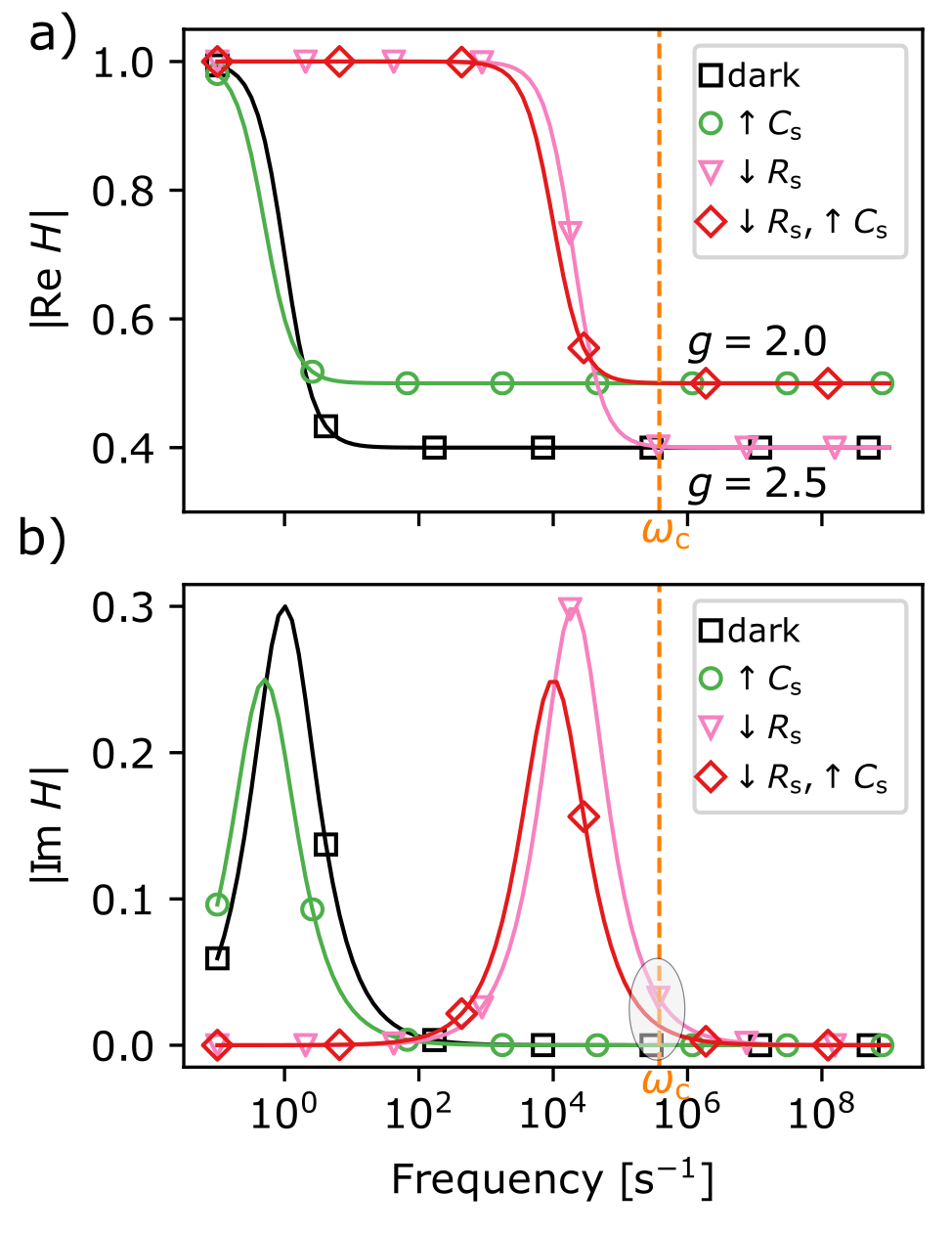}
\caption[Transfer function]{(a) Real and (b) imaginary part of the Eq.~\ref{eq:H-full} transfer function plotted for the following cases: 
(i) sample in the dark and under illumination with 
(ii) increased sample capacitance ($\uparrow C\st{s}$), 
(iii) decreased sample resistance ($\downarrow R\st{s}$), and 
(iv) increased sample capacitance and decreased sample resistance ($\downarrow R\st{s}$, $\uparrow C\st{s}$). 
To mimic an increase in $C\st{s}$, the value of $g$ in Eq.~\ref{eq:H-full} was decreased from $2.5$ to $2.0$. 
The cantilever frequency $\omega\st{c}$ is indicated as a dashed vertical line.
The sample-induced frequency shift is proportional to ${\mathrm{Re}} \, H(\omega\st{c})$, (a), while sample-induced dissipation is proportional to ${\mathrm{Im}} \, H(\omega\st{c})$, (b).
The gray zone highlights the region in (b) that determines ${\mathrm{Im}} \, H(\omega\st{c})$.}
\label{fig:transfer-function-cases}
\end{figure}

These conclusions can be visualized with the help of the simulated spectra shown in Fig.~\ref{fig:transfer-function-cases}a.
As explained in Refs.~\citenum{Tirmzi2017jan}, \citenum{Tirmzi2019jan}, and \citenum{Dwyer2019jun}, the BLDS signal $\alpha$ is proportional to $|H(\omega\st{m})|^2$, which is dominated by $\Re{\! \big[H(\omega\st{m})\big]}$ at high frequency.
In Fig.~\ref{fig:transfer-function-cases}a we plot representative $\Re{\! \big[H(\omega\st{m})\big]}$ versus $\omega\st{m}$ spectra in the dark and under three light-on scenarios: increased $C\st{s}$, decreased $R\st{s}$, and both changing together.
Only the third scenario is qualitatively consistent with the Fig.~\ref{fig:frequency-BLDS} spectra. 

These conclusions about photoinduced changes in sample capacitance and resistance are corroborated by the dissipation data of Fig.~\ref{fig:dissipation-ringdown}.
The dissipation signal is proportional to \cite{Tirmzi2017jan,Tirmzi2019jan,Dwyer2019jun} $\Im{\! \big[H(\omega\st{c})\big]}$, the imaginary part of the transfer function $H$ evaluated at the cantilever frequency $\omega\st{c}$.
In Fig.~\ref{fig:transfer-function-cases}b we plot representative $\Im{\! \big[H\big]}$ spectra for the four scenarios, with the cantilever frequency $\omega\st{c} = 2 \pi \, f\st{c}$ indicated by a dotted vertical line and the value of the transfer function near $\omega\st{c}$ indicated by gray shading. 
Here either the second or third light-on scenario are consistent with the observed increase in dissipation with increasing light intensity.
The observed dissipation unconditionally requires a decrease in $R\st{s}$ with increasing light intensity.

We note that the numerical value of the time constant $\omega\st{fast}^{-1}$ deduced from the BLDS spectra will depend on both sample capacitance and tip capacitance and this value may not represent the ``life time'' of carrier generation or recombination in the sample.
Nevertheless, changes in this time constant can be directly related to changes in sample conductivity.

\section{tr-EFM and pk-EFM}
Frequency shift (\textit{e.g.} tr-EFM and FF-tr-EFM) \cite{Coffey2006sep,Giridharagopal2012jan,Karatay2016may} and phase shift methods (pk-EFM) \cite{Dwyer2017jun,Dwyer2019jun} have been the recent focus of high temporal resolution SPM measurements and have shown the ability to achieve time resolution down to $1 \%$ of the cantilever cycle \cite{Dwyer2017jun}.
To date, the origin of the observed frequency shift and the accompanying phase shift has been attributed solely to changes in sample capacitance \cite{Karatay2016may, Dwyer2017jun}.
Here we show that light-induced changes in sample conductivity better explain the data in samples having an appreciable impedance, such as our lead halide perovskites.

\begin{figure*}[t]
\includegraphics[width=6in]{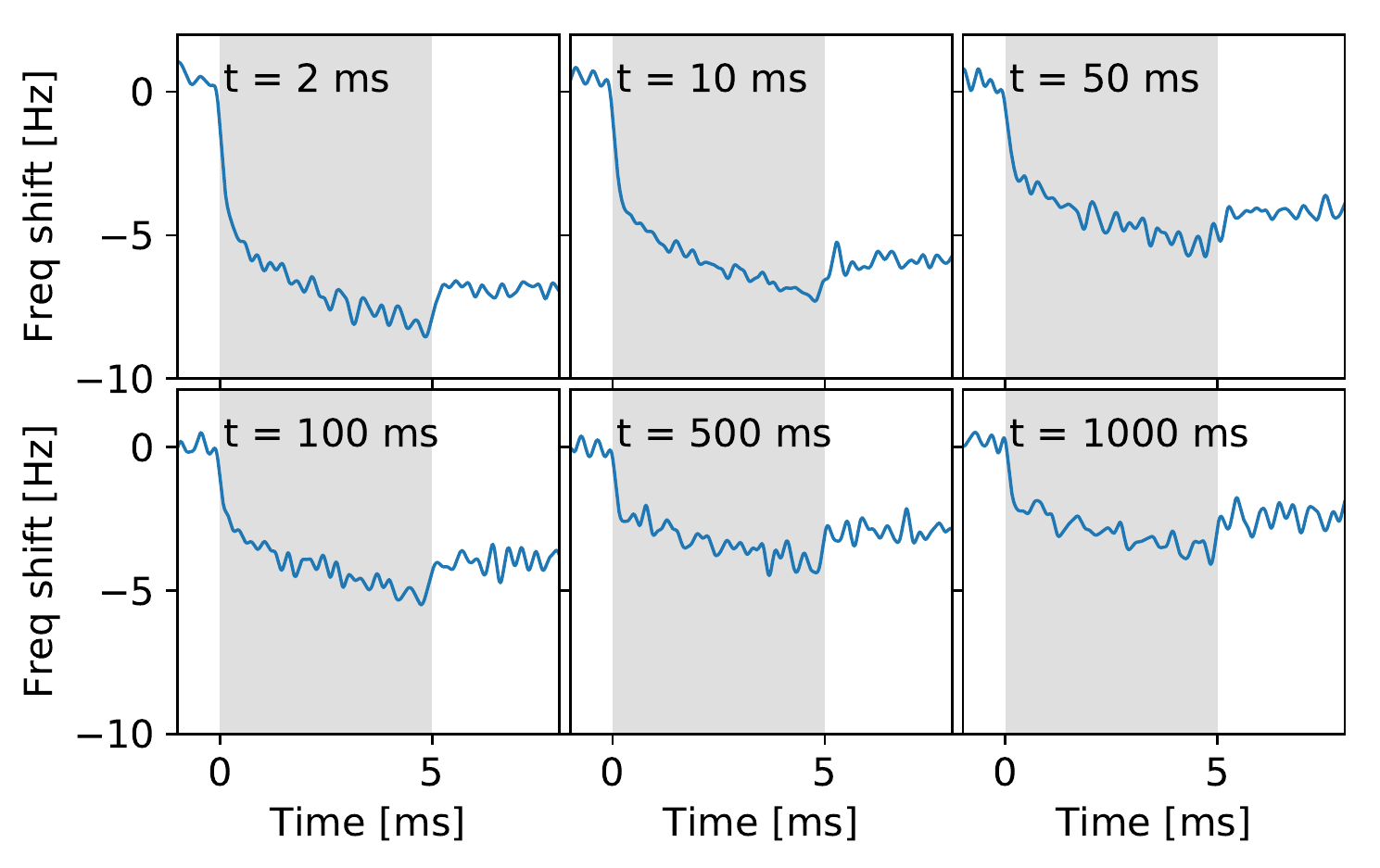}
\caption[tr-EFM]{The light-induced frequency shift is determined by the sample bias time $t$ in the dark.
tr-EFM traces with different bias time before the illumination period (shaded region).
See Fig.~\ref{fig:tr-EFM-t} for the associated experimental timing diagram. 
Experimental parameters: $I_{h\nu} = \SI{120}{\milli\W\per\cm\squared}$, $V\st{ts} = \SI{6}{\V}$, $h = \SI{200}{\nano\meter}$, and $\lambda = \SI{405}{\nano\meter}$.}
\label{fig:trEFM-tp-time}
\end{figure*}

We performed  tr-EFM experiments in which the sample was biased for a time $t$ in the dark before light was applied.
In Fig.~\ref{fig:trEFM-tp-time} we show tr-EFM traces of cantilever frequency versus time for different values of the delay time ($t$) between the start of the voltage pulse and the start of the light pulse. 
For a short delay $t$ such as $\SI{2}{\milli\second}$, we see an $\SI{8}{\hertz}$ light-induced frequency shift with a fast component happening on the $\approx$ $\SI{100}{\micro\second}$ time-scale and a slow component evolving on the millisecond timescale.
With increasing delay $t$, both the fast and slow components decrease in magnitude so that for $t=\SI{1000}{\ms}$, the light-induced frequency shift is \SI{4}{\Hz} (observation 1).
When the light is turned off at \SI{5}{\ms}, most of the light-induced frequency shift is retained (observation 2). 
In Fig.~\ref{fig:tr-EFM-1sec}, we show the full trace of the tr-EFM measurement for $t =  \SI{1}{\second}$.
These two key observations cannot be easily explained in terms of a photocapacitance, since photocapacitance should not depend on the delay $t$ and should revert back to the original dark capacitance when the light is turned off.

What is the physical explanation for the finding that sample bias in the dark strongly affects the subsequent light-induced frequency shift?
The key insight from Eq.~\ref{eq:frequency_shift_time_varying} is that the frequency shift depends on both $\Re{[H(\omega\st{c}, t)]}$, the time-varying transfer function evaluated at the cantilever resonance frequency, and the time-dependent tip charge $q\st{t}$ or tip voltage $V\st{t} = q\st{t}/C\st{tip}$.
From Fig.~\ref{fig:frequency-BLDS}, we know $\Re{[H(\omega\st{c})]}$ increases with light because the sample conductivity increases.
This increase in sample conductivity explains the two key observations from Fig.~\ref{fig:trEFM-tp-time} once we properly recognize the role of the tip voltage drop $V\st{t}$.

\begin{figure*}
\includegraphics[width=6.5in]{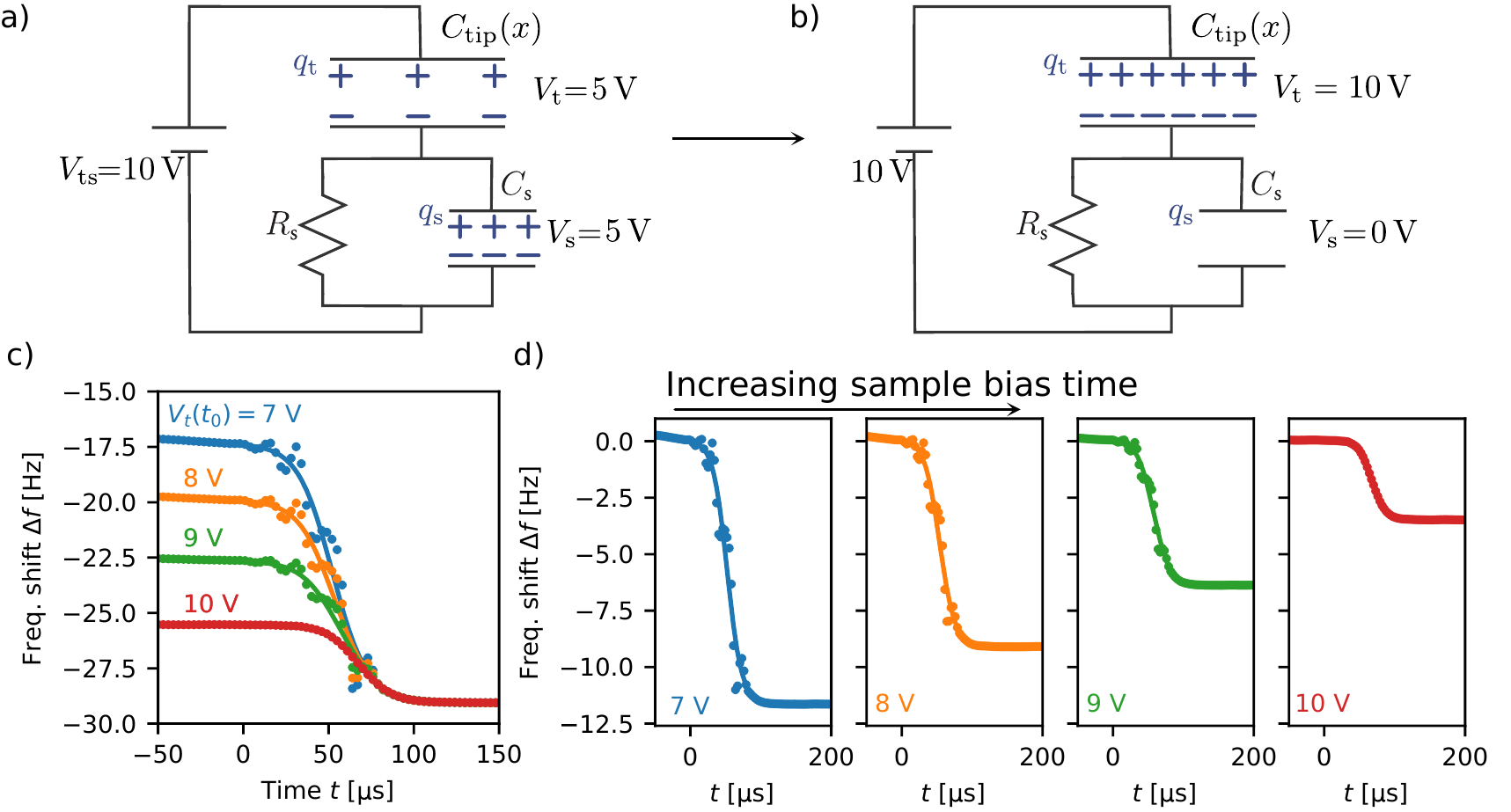}
\caption[The effect of the initial tip voltage on light-induced frequency shift.]{The effect of the initial tip voltage on frequency-shift transients. 
(a) The model circuit used in this simulation. After the tip-sample voltage $V\st{ts}$ is applied, charge builds up on the tip capacitor and sample capacitor.
(b) Eventually, the sample capacitor discharges and $V\st{t} = V\st{ts}$.
(c) Frequency shift versus time calculated by numerical simulation (points) and the Eq.~\ref{eq:frequency_shift_time_varying} analytical approximation  (lines) for different initial tip voltages $V\st{t}(t_0) = q\st{t}^{(0)}(t_0)/C\st{tip} = 7, 8, 9, \SI{10}{\V}$. 
(d) The data of (c), shifted so $\Delta f(t=0) = 0$ for each trace to more easily compare with Fig.~\ref{fig:trEFM-tp-time}.
All traces were smoothed by averaging the frequency shift over a single period twice.
Simulation parameters:
$R\st{dark} =$ \SI{1e7}{\mega\ohm};
$R\st{light} =$ \SI{1e4}{\mega\ohm}; 
$\tau\st{L} =$ \SI{10}{\us}; 
$V\st{t}(t_0) = 7$, $8$, $9$ and $\SI{10}{\V}$ (blue, orange, green, red);
$C\st{s} = $ \SI{0.05}{\fF};
$C\st{tip} = $ \SI{0.1}{\fF};
$C\st{tip}' = $ \SI{-28}{\fF\per\nm};
$f\st{c} = \SI{62.5}{\kHz}$;
$k\st{c} = \SI{3.5}{\N\per\m}$;
$Q = \num{26000}$, $A_0 = \SI{50}{\nm}$;
$A_0 = \SI{50}{\nm}$;
$t_0 = \SI{-200}{\us}$;
$\phi_0 = 0$ (see SI Section~\ref{trEFMandpkEFM}).
}
\label{fig:compare-Vt}
\end{figure*}

In the Fig.~\ref{fig:trEFM-tp-time} tr-EFM experiment, a tip-sample voltage on the order of $5$ to $\SI{10}{\volt}$ was applied before the light was turned on.
The circuit schematics in Fig.~\ref{fig:compare-Vt}(a,b) show the modeled tip voltage $V\st{t}(t)$ when $C\st{s} = C\st{tip}$ ($g =2$).
When the tip-sample bias is applied, equal charges build up on both capacitors and the voltage drop across the tip capacitor is $V\st{t} = V\st{ts} C\st{s}/(C\st{s} + C\st{tip})$ (Fig.~\ref{fig:compare-Vt}a).
The sample capacitor discharges with an $RC$ time constant $R\st{s}(C\st{tip} + C\st{s})$ until eventually all of the applied tip-sample voltage drops across the tip capacitor: $V\st{t} = V\st{ts}$ (Fig.~\ref{fig:compare-Vt}b).
In the dark the sample is very resistive so it can take seconds for this equilibrium condition to be reached.
If the light is turned on before the tip is fully charged, the increased sample conductivity (decrease in sample resistance $R\st{s}$) causes the tip to charge more quickly, increasing $V\st{t}$ and therefore causing a larger light-induced frequency shift.

Figure~\ref{fig:compare-Vt}(c,d) shows a simulation to illustrate how this tip charging explains the data of Fig.~\ref{fig:trEFM-tp-time}.
It was assumed the sample resistance $R\st{s}$ decreases when the light is turned on.
The evolution of tip position $x$ and tip and sample charge are described by a set of coupled differential equations (see SI Section~\ref{trEFMandpkEFM} and Fig.~\ref{fig:df_t_explained} for details).
Fig.~\ref{fig:compare-Vt}(c,d) shows the frequency shift calculated numerically from $x(t)$ (points) and analytically from Eq.~\ref{eq:frequency_shift_time_varying} (curves).
The four traces show increasing initial tip voltages that correspond to the increasing bias times in Fig.~\ref{fig:trEFM-tp-time}.
With reasonable values of sample resistance in the dark, sample resistance in the light, sample capacitance, and tip capacitance, the numerical simulations and Eq.~\ref{eq:frequency_shift_time_varying} do an excellent job qualitatively reproducing observation 1; the closer the initial tip voltage $V\st{t}$ is to its steady-state value $V\st{t} = V\st{ts} = \SI{10}{\V}$, the smaller the magnitude of the observed frequency shift. 
The slow component of the frequency shift in Fig.~\ref{fig:trEFM-tp-time} can be explained by the millisecond-scale $RC$ time apparent in the BLDS spectra under steady-state illumination.
To corroborate this hypothesis, we tracked changes in $\alpha$ at $\omega\st{m} = \SI{1257}{\per\second}$ ($1/\omega\st{m} = \SI{0.8}{\milli\second}$) in real time when the light was turned on (Fig.~\ref{fig:fixed-freq-SI}; temporal resolution is limited to $\SI{1}{\second}$ by the finite amplitude-modulation frequency and associated lock-in measurement).
The dielectric response changes in $\leq \SI{1}{\second}$, consistent with the timescales seen in Fig.~\ref{fig:trEFM-tp-time}.

The retention of light-induced frequency shift in the dark (observation 2) can also be explained by carefully considering the evolution of the tip voltage $V\st{t}$.
When the light is turned off, the tip does not discharge, since $V\st{t} = V\st{ts}$ is the equilibrium state in both the light and dark.
Therefore any frequency shift caused by an increase in $V\st{t}$ is retained even after the light is turned off.
This explains why the change in $\Delta f$ when the light is turned off in Fig.~\ref{fig:trEFM-tp-time} is relatively small no matter how long the initial pre-bias time was.
Any decrease in frequency shift when the light is turned off is attributed to changes in $\Re{H(\omega\st{c}, t)}$ (Eq.~\ref{eq:frequency_shift_time_varying}), since only changes in $\Re{H(\omega\st{c}, t)}$ are reversed when the light turns off.

\begin{figure}[t]
\includegraphics[width=3.25in]{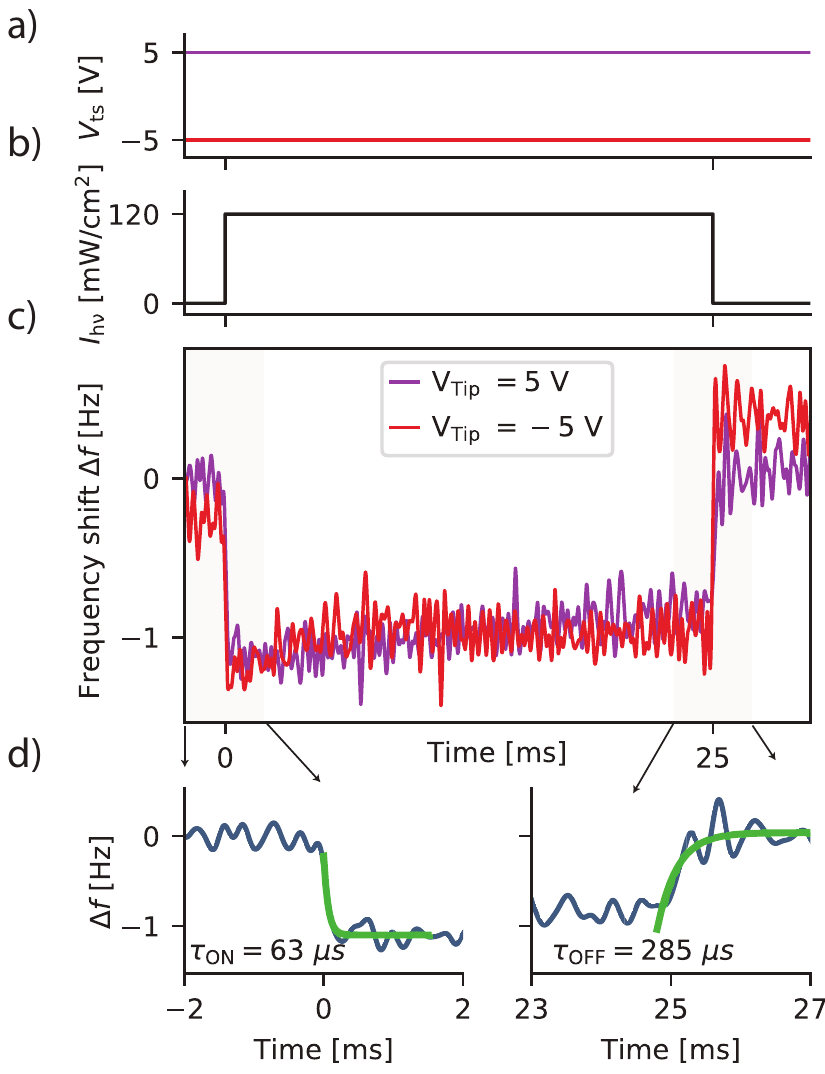}
\caption[tr-EFM]{Light-induced changes in cantilever frequency likewise occur promptly.
tr-EFM frequency shift signals with a DC bias of $\pm$ $\SI{5}{\volt}$. 
(a) The tip-sample voltage is applied continuously.
(b) Light is switched on at time $t = \SI{0}{\milli\second}$ and switched off at $t = \SI{25}{\milli\second}$. 
(c) Light induced frequency shift measured for $V\st{ts}$ $=$ $\pm$ $\SI{5}{\volt}$.
(d) Zoom-in of the $+\SI{5}{\volt}$ trace.
Solid (green) lines are a fit to a single exponential, with light-on and light-off time constants indicated.
Experimental parameters:  $I_{h\nu} = \SI{120}{\milli\W\per\cm\squared}$, $h = \SI{200}{\nano\meter}$, $\lambda = \SI{405}{\nano\meter}$.}
\label{fig:trEFM-tp-constant}
\end{figure}

The SI Section~\ref{trEFMandpkEFM} analysis reveals that the frequency-shift dynamics are greatly simplified if the tip charge, \emph{i.e.}\ $V\st{t}$, has reached steady state before light is applied.
We can achieve the steady-state condition by applying a DC voltage for an extended period of time before beginning the measurement (Fig.~\ref{fig:trEFM-tp-constant}).
Pre-biasing the sample with constant applied voltage for 5 minutes before the measurement and keeping the bias always on during the measurement keeps the tip charged ($V\st{t} = V\st{ts}$); therefore, all observed frequency shifts reflect changes in the transfer function at the cantilever frequency $\Re{[H(\omega\st{c}, t)]}$. 

For $V\st{ts} = \SI{5}{V}$ and $\SI{-5}{V}$, the light induced frequency changes are essentially identical, with similar rise and fall time on the order of $\approx$ $\SI{100}{\micro\second}$.
The insensitivity of the frequency shift, rise time, and fall time to the sign of the applied tip-sample voltage rules out any effects on the measurement due to surface potential changes.
When the light is turned off at $t=\SI{25}{\ms}$, the frequency shift reverts to the dark value, unlike in Fig.~\ref{fig:trEFM-tp-time}.
This observation is consistent with Eq.~\ref{eq:frequency_shift_time_varying} when $V\st{t}$ is constant---the light-induced changes in $\Re{[H(\omega\st{c})]}$ are fully reversible.
Considered together, the data of Fig.~\ref{fig:dissipation-ringdown}b and Fig.~\ref{fig:trEFM-tp-constant}d allow us to conclude that the transfer function at $\omega\st{c}$, determining the light-induced frequency shift and dissipation, is evolving on the $\SI{100}{\micro\second}$ timescale.

\begin{figure}[t]
\includegraphics[width=3.25in]{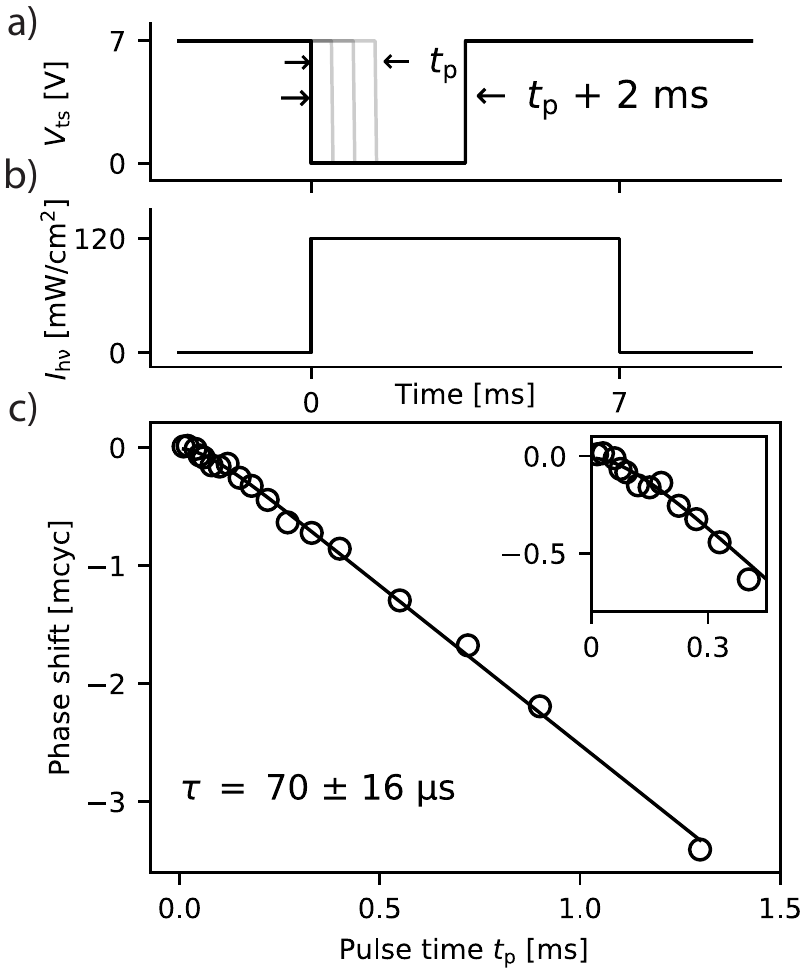}
\caption[pkEFM]{pK-EFM measures a photoconductivity rise time of $\tau\st{on} = \SI{70}{\micro\second}$. 
Timing diagram of applied (a) tip voltage and 
(b) light, 
with $t\st{p}$ the duration of voltage- and light-pulse overlap.
(c) Measured cantilever phase shift $\Delta \phi$ versus pulse time $t\st{p}$.  
Inset: Short $t\st{p}$ data.
The solid line is a fit to $\Delta \phi = \Delta \omega\st{ss} (t\st{p} - \tau\st{on} + \tau\st{on} e^{-t\st{p}/\tau\st{on}})$, Eq.~19 in Ref.~\citenum{Dwyer2017jun}, with $\Delta \omega\st{ss}$ the steady-state light-induced cantilever frequency shift and $\tau\st{on} = \SI{70}{} \pm \SI{16}{\micro\second}$ the photoconductivity rise time.
Experimental parameters: $I_{h\nu} = \SI{120}{\milli\W\per\cm\squared}$,  $V\st{ts} = \SI{7}{\V}$, $h = \SI{200}{\nano\meter}$, $\lambda = \SI{405}{\nano\meter}$.}
\label{fig:pkEFM-tp-constant}
\end{figure}

To better resolve this time constant, we carried out a pk-EFM experiment in which the applied tip-sample voltage is applied continuously except during a $\sim \! \! \SI{2}{\milli\second}$ time period before the end of light pulse.
This short time without applied bias is necessary to measure cantilever phase changes during the pulse time $t\st{p}$ arising from time-dependent sample conductivity.
From a plot of the measured phase shift versus pulse time, we deduce a time-constant of $\SI{70 \pm 16}{\micro\second}$.

\section{Discussion}
The steady-state impedance spectra and photoconductivity dynamics of the 2D lead-halid perovskite \ce{(C4H9NH3)2PbI4} observed using electrical scanned-probe measurements are qualitatively different from those found in the 3D lead-halide perovskite \ce{CsBrBr3} \cite{Tirmzi2017jan} and the state-of-the-art FAMACs alloy \cite{Tirmzi2019jan}.

In Ref.~\citenum{Tirmzi2019jan} we used Maier's transmission-line treatment of impedance spectroscopy \cite{Jamnik1999apr,Lai2005nov} to show that in a mixed electronic-ionic conductor, for reasonable assumptions about size of the electronic and ionic conductivities, the scanned-probe broadband local dielectric spectroscopy (BLDS) experiment of Fig.~\ref{fig:frequency-BLDS} measures the \emph{total} sample conductivity.
Modeling the sample as a parallel resistor-capacitor circuit and assuming $C\st{tip} \geq C\st{s}$, which is approximately valid here, the associated rolloff frequency depends on sample resistance and is largely independent of sample capacitance.
In this limit, the rolloff frequency is given by $\omega\st{fast} \approx A \sigma\st{s} \big/ L C\st{tip}$ with $A$ the sample area under the tip, $L$ the sample thickness, $C\st{tip}$ the tip-sample capacitance, and $\sigma\st{s} = \sigma\st{eon} + \sigma\st{ion}$ the electronic plus ionic conductivity of the sample.
Since values of $A$, $L$, and $C\st{tip}$ used in the Fig.~\ref{fig:frequency-BLDS} BLDS experiment on \ce{(C4H9NH3)2PbI4} are similar to values used in related experiments on \ce{CsPbBr3} \cite{Tirmzi2017jan}, $\sigma\st{s} \propto\omega\st{fast}$, and we can draw qualitative conclusions about the samples' conductivity by comparing the measured BLDS spectra and rolloff frequency in the two samples. 

In \ce{CsPbBr3}, the impedance spectrum showed two distinct rolloffs \cite{Tirmzi2017jan} --- one below \SI{1}{\hertz}, attributed in the dielectric spectroscopy literature to ionic motion, and a second rolloff that varied from \SI{1E1}{\hertz} in the dark to above \SI{1e5}{\hertz} at \SI{1000}{\milli\watt\per\square\centi\meter}.
The second rolloff frequency, and hence the associated $\sigma\st{s}$, had a nearly square root dependence on illumination intensity, as one would expect for free carriers, and yet the conductivity exhibited a slow, activated recovery consistent with ion or vacancy motion.
We subsequently proposed \cite{Tirmzi2019jan} that these contradictory observations could be rationalized by considering that photoexcitation can create interstitial iodine atoms and mobile iodine vacancies \cite{Kim2018mar}: in Kr\"{o}ger-Vink notation, $\text{I}_{\st{I}}^{\text{x}}
  \xrightarrow{\: \: h \nu \: \:}
  \text{I}_{\st{I}}^{\text{x}} 
  + \text{h}^{\bullet}
  + \text{e}^{\prime}
  \ce{<=>} 
  \text{I}_{\st{i}}^{\text{x}} 
  + \text{V}_{\st{I}}^{\bullet}
  + \text{e}^{\prime}$.
The scanned-probe BLDS experiment measures the total conductivity and would therefore be sensitive to both $\text{V}_{\st{I}}^{\bullet}$ and $\text{e}^{\prime}$.
In this view, the \emph{circa} \SI{10}{\second} to \SI{100}{\second} conductivity recovery observed in \ce{CsPbBr3} \cite{Tirmzi2017jan} and the FAMACs alloy \cite{Tirmzi2019jan} is measuring the slow, activated return to equilibrium of photogenerated halide vacancies and their geminate electrons. 
  
In \ce{(C4H9NH3)2PbI4}, in contrast, only one rolloff was apparent in the BLDS spectrum and the associated cutoff frequency varied from $\leq \SI{1}{\hertz}$ in the dark to \SI{e3}{\hertz} at \SI{1000}{\milli\watt\per\square\centi\meter}.
Compared to the \ce{CsBrBr3} film, the total conductivity of \ce{(C4H9NH3)2PbI4} is 100-fold smaller under illumination and saturates at high illumination intensity. 
We can make a rough estimate of the sample's dark conductivity using $\sigma\st{s} = L C\st{tip} \omega\st{fast} \big/ A$ with $\omega\st{fast} = \SI{1}{\hertz}$. 
The sample thickness is $L = \SI{500}{\nano\meter}$ and we estimate $C\st{tip} = \SI{1e-16}{F}$ and $A = \pi (h/2)^2$, with $h = \SI{150}{\nano\meter}$ the tip-sample separation.
Using these parameters, we obtain $\sigma\st{s} = \SI{4e-13}{\siemens\per\centi\meter}$, in reasonable agreement with the dark conductivity of $\SI{1e-13}{\siemens\per\centi\meter}$ measured for a single crystal of the 2D perovskite \ce{(PEA)2PbI4} \cite{Peng2017jun}.

In \ce{(C4H9NH3)2PbI4} the photoconductivity rise time $\tau\st{on}$ (Figs.~\ref{fig:trEFM-tp-constant} and \ref{fig:pkEFM-tp-constant}) and decay time $\tau\st{off}$ (Fig.~\ref{fig:trEFM-tp-constant}) are \SI{70}{\micro\second} and \SI{285}{\micro\second}, respectively.
The $\tau\st{on}$ and $\tau\st{off}$ times measured here are $\sim$\SI{e5}{} faster than the conductivity decay times seen in multiple 3D lead-halide perovskites \cite{Tirmzi2017jan,Tirmzi2019jan} but $\sim$\SI{e4}{} slower than the \SI{10}{\nano\second} electron-hole recombination times observed for \ce{(C4H9NH3)2PbI4} using time-resolved microwave conductivity \cite{Venkatesan2018jan}.
The $\tau\st{on} = \SI{70\pm16}{\micro\second}$ measured here using pk-EFM is in good agreement with the cantilever-frequency equilibration time of \SI{85}{\micro\second} measured near grain centers in \ce{(C4H9NH3)2PbI4} by Giridharagopal, Ginger, and coworkers using FF-tr-EFM \cite{Giridharagopal2019feb}.

In Ref.~\citenum{Giridharagopal2019feb}, spatial variations in signal were discussed in terms of ionic and electronic carrier transport, yet in the usual description of tr-EFM the frequency-shift signal, expressed in terms of a time-dependent tip-sample capacitance, has no overt connection to such transport processes.
Building on prior work, here we provide that missing theoretical connection.
In Ref.~\citenum{Dwyer2019jun} we showed, for a sample described by a parallel resistor-capacitor circuit, that the frequency shift and dissipation probe the real and imaginary parts, respectively, of the Eq.~\ref{eq:H-full} transfer function $H(\omega)$.
The BLDS spectra of Fig.~\ref{fig:frequency-BLDS} and the qualitative changes in $H(\omega)$, frequency shift, and dissipation sketched in Fig.~\ref{fig:transfer-function-cases} reveal that the frequency shift observed by FF-tr-EFM and pk-EFM in \ce{(C4H9NH3)2PbI4} is dominated by sample resistance, not capacitance.
Motivated by this observation, here we extended our Ref.~\citenum{Dwyer2019jun} treatment of transient electric force microscope experiments to include a time-dependent sample resistance (\emph{i.e.}, conductivity).
This finding shows that pk-EFM data can in principle be directly compared to time-dependent microwave conductivity measurements.
Since pk-EFM (and FF-tr-EFM) is sensitive to \emph{both} time dependent capacitance and resistance, future time-resolved EFM studies should include measurements of BLDS spectra versus light intensity in order to establish whether $C\st{s}$ or $R\st{s}$ is dominating the sample's response.  
Because $ \omega\st{fast} \ll \omega\st{c}$ here, the sample's changing conductivity has only a small effect on the cantilever frequency and dissipation here; imaging spatial variations in \ce{(C4H9NH3)2PbI4}'s steady-state photoconductivity would be better accomplished using a BLDS signal ($\omega\st{m} \sim \SI{100}{\hertz}$ to \SI{1000}{\hertz}) than a frequency or dissipation signal.

While the electronic and ionic conductivity of \ce{(C6H10N2)PbI4} in the dark has been measured previously \cite{Lermer2018aug}, there is little precedent beyond Ref.~\citenum{Giridharagopal2019feb} for observing a \emph{circa} \SI{100}{\micro\second} conductivity rise time and decay time following optical irradiation for a 2D lead-halide perovskite.  
There are two general explanations for such a conductivity-relaxation time.
For photogenerated electronic carriers to equilibrate this slowly would require trapping; hole trapping at acceptor-type iodine vacancies is plausible \cite{Peng2017jun}, as is localization at iodine edge states \cite{Zhang2019sep}.
In the case of ionic conductivity, the rise and decay could be due to a dependence of ionic conductivity on the concentration of (decaying, trapped) electrons and holes.
Alternatively, we could be observing the formation and relaxation of photogenerated vacancies; in this case we would have to explain why the formation and relaxation rate is \SI{e5}{} times faster than in 3D perovskites.
In support of this conjecture, Wang \emph{et al.} found the light-on and light-off transient time for a single-crystal photodetector of BAPI to be \SI{e3}{} larger than for a 3D perovskite \cite{Wang2018apr}.  
The fast rise time is puzzling, because we would expect the vacancy formation energy to be high in our 2D perovskite based on the comparatively low dark conductivity seen in related systems \cite{Peng2017jun}.
Fast relaxation is more feasible because of the reduced out-of-plane ion motion expected from the insulating organic layers; the associate barrier to diffusion would keep the product interstitial halide physically close to the vacancy and ready for the back reaction.
Considering all these observations, we see little clear evidence of light-induced vacancy formation in BAPI, in contrast with the 3D perovskites \cite{Tirmzi2017jan,Tirmzi2019jan}.


\section{Conclusion}

We have shown that the transfer function representation of SPM used to describe broadband local dielectric spectroscopy and dissipation measurements can be extended to write the temporal response measured in tr-EFM and pk-EFM in terms of sample resistance $R\st{s}$ \cite{Dwyer2019jun}.
We have established experimentally using a range of steady-state and time-resolved studies that electrical scanned probe measurements in BAPI are primarily observing changes in sample resistance and not capacitance. 
We find, surprisingly, that light-induced conductivity in the 2D BAPI system recovers \SI{e5}{} faster than in many 3D perovskite systems studied to date \cite{Tirmzi2017jan,Tirmzi2019jan}.


\bigskip

\noindent {\large \textbf{SUPPORTING INFORMATION AVAILABLE}} 
The Supporting Information contains: theoretical derivation for Eq.~\ref{eq:frequency_shift_time_varying}; numerical simulations for frequency and phase shift in tr-EFM and pk-EFM for variety of experimental conditions \cite{Dwyer2020may_data}; timing diagram for Fig.~\ref{fig:trEFM-tp-time}; dielectric response at $\omega\st{m}$ $=$ $\SI{1257}{\per\second}$; absorbance spectra; XRD; AFM topography image; surface potential at selected light intensities.


\bigskip

\noindent {\large \textbf{AUTHOR INFORMATION}}

\bigskip


\medskip

\noindent \textbf{Corresponding Author}

\medskip

\noindent ${}^{*}$E-mail: \href{mailto:jam99@cornell.edu}{jam99@cornell.edu}

\noindent Faculty webpage: \\
\href{http://chemistry.cornell.edu/john-marohn}{http://chemistry.cornell.edu/john-marohn}

\noindent Research group webpage: \\
\href{http://marohn.chem.cornell.edu/}{http://marohn.chem.cornell.edu/}

\medskip


\medskip

\medskip

\noindent \textbf{Notes}

\medskip

\noindent The authors declare no competing financial interest.

\medskip

\begin{acknowledgement}
A.M.T and J.A.M acknowledge the financial support of the U.S. National Science Foundation (Grant
DMR-1709879).
F.J. acknowledges support from the U.S. Department of Energy, Office of Basic Energy Sciences, grant DE-SC0013957.
The authors gratefully acknowledge David S. Ginger for thoughtful discussions.  

\end{acknowledgement}

\providecommand{\latin}[1]{#1}
\makeatletter
\providecommand{\doi}
  {\begingroup\let\do\@makeother\dospecials
  \catcode`\{=1 \catcode`\}=2 \doi@aux}
\providecommand{\doi@aux}[1]{\endgroup\texttt{#1}}
\makeatother
\providecommand*\mcitethebibliography{\thebibliography}
\csname @ifundefined\endcsname{endmcitethebibliography}
  {\let\endmcitethebibliography\endthebibliography}{}

\label{TheEnd}
\end{document}


\title{Supporting Information: Light-dependent Impedance Spectra and Transient Photoconductivity in a Ruddlesden--Popper 2D Lead-halide Perovskite Revealed by Electrical Scanned Probe Microscopy and Accompanying Theory}
    
	\author{Ali Moeed Tirmzi}
	\affiliation{Dept.\ of Chemistry and Chemical Biology, Cornell University, Ithaca, NY 14853, USA}
	\author{Ryan P. Dwyer}
	\affiliation{Dept.\ of Chemistry and Biochemistry, University of Mount Union, Alliance, OH 44601}
	\author{Fangyuan Jiang}
        \affiliation{Department of Chemistry, University of Washington, Seattle, Washington 98195, United States}
	\author{John A. Marohn}
	\affiliation{Dept.\ of Chemistry and Chemical Biology, Cornell University, Ithaca, NY 14853, USA}
	\email{jam99@cornell.edu}

\maketitle
\thispagestyle{empty}

\section{Theory: Derivation of Eq.~\ref{eq:frequency_shift_time_varying}}
\label{sec:Theory-derivation}
To derive Eq.~\ref{eq:frequency_shift_time_varying}, we start from the equation for the phase shift $\Delta \phi$ induced by the tip-sample force $F\st{ts}$ \cite{Giessibl1997dec},
\begin{equation}
\Delta \phi = - \frac{f\st{c}}{k\st{c} A_0^2} \int_0^t F\st{ts}(t') x(t') dt',
\label{eq:Delta_phi1}
\end{equation}
where $f\st{c}$ is the cantilever resonance frequency, $k\st{c}$ is the cantilever spring constant, $A_0$ is the cantilever's initial amplitude at $t=0$, and $x(t)$ is the cantilever's displacement versus time. 
Following the theory introduced in Ref.~\citenum{Dwyer2019jun}, the tip-sample force, tip-sample charge, and cantilever displacement are approximated using perturbation theory.
We use the notation $x^{(0)}$ to represent the zeroth-order approximation for $x(t)$.
From Ref.~\citenum{Dwyer2019jun}, the zeroth order tip-sample force is 
\[
F\st{ts}^{(0)}(t) = \frac{1}{2}C\st{tip}'  \frac{\left [ q\st{t}^{(0)}(t)\right ]^2}{C\st{tip}^2}= \frac{1}{2}C\st{tip}' V\st{t}(t)^{2}
\]
where $q^{(0)}(t)$ is the zeroth order tip voltage and $V\st{t}$ is the zeroth order tip voltage
\begin{equation}
V\st{t} = \frac{q\st{t}^{(0)}(t)}{C\st{tip}}.
\label{eq:Vt-def}
\end{equation}
The zeroth order tip-sample force only causes a phase shift if $V\st{t}^2$ contains significant content at the cantilever frequency $\omega\st{c}$. For now, we assume that $V\st{t}^2$ varies slowly enough that it does not cause a phase shift.

The first-order tip-sample force $F\st{ts}^{(1)}$ describes the oscillating forces caused by the oscillating tip. These oscillating forces cause the frequency shift in most KPFM experiments. In the perturbation theory approximation, the first-order tip-sample force is 
\begin{equation}
F\st{ts}^{(1)} = \frac{C''_q q\st{t}^{(0)} q\st{t}^{(0)}x^{(0)}}{2 C\st{tip}^2}
+  \frac{C\st{tip}' q\st{t}^{(0)} q\st{t}^{(1)}}{C\st{tip}^2}
\label{eq:Fts_1}
\end{equation}
where the capacitance derivative $C''_q$ is
\begin{equation}
C''_q = C\st{tip}'' - 2(C\st{tip}')^2 / C\st{tip}.
\label{eq:Czz_q_def}
\end{equation}
The two terms of Eq.~\ref{eq:Fts_1} describe the two possible causes of the oscillating tip-sample force. In the first term, the cause is oscillations in the tip-sample energy arising from the oscillating tip displacement (at constant charge). In the second term, the cause is the oscillating charge that flows in response to the oscillating tip displacement.
The resulting phase shift is obtained by substituting back into Eq.~\ref{eq:Delta_phi1} 
\begin{equation*}
\Delta \phi = - \frac{f\st{c}}{k\st{c} A_0^2} \int_0^t \left [\frac{C''_q q\st{t}^{(0)} q\st{t}^{(0)}x^{(0)}}{2 C\st{tip}^2}
+  \frac{C\st{tip}' q\st{t}^{(0)} q\st{t}^{(1)}}{C\st{tip}^2} \right ]  x(t') dt'.
\end{equation*}
We approximate $x(t)$ by its zeroth-order approximation $x^{(0)}(t)$. In these experiments, the cantilever is excited at its resonance frequency so $x^{(0)}(t) = A_0 \cos( \omega\st{c} t + \phi_0)$. For simplicity, we take the initial cantilever phase to be $\phi_0 = 0$; none of the conclusions below depend on this assumption.
After plugging in for $x(t)$, the phase shift simplifies to
\begin{equation}
\Delta \phi = \underbrace{- \frac{f\st{c}}{k\st{c}} \int_0^t \frac{C''_q q\st{t}^{(0)} q\st{t}^{(0)}}{2 C\st{tip}^2} \cos^2(\omega\st{c} t') dt'
 }_{\text{constant charge term}} \, \, \underbrace{- \frac{f\st{c}}{k\st{c} A_0}\int_{0}^{t}\frac{C\st{tip}' q\st{t}^{(0)} q\st{t}^{(1)}}{C\st{tip}^2} \cos(\omega\st{c} t')   dt'}_{\text{oscillating charge term}}.
\label{eq:Delta_phi_Fts1}
\end{equation}

To determine the cantilever phase shift during these experiments, we consider the two terms in Eq.~\ref{eq:Delta_phi_Fts1} individually. The first term ($\Delta \phi\st{const}$) describes the phase shift caused by the tip oscillating at constant charge. The second term ($\Delta \phi\st{osc}$) describes an additional phase shift caused by the tip charge oscillating along with the oscillating tip displacement. 
Using the definition of $V\st{t}$ (Eq.~\ref{eq:Vt-def}) and the trigonometric identity for $\cos^2\theta$, the first term simplifies to
\begin{equation}
\Delta \phi\st{const} = - \frac{f\st{c}}{4 k\st{c}} C''_q \int_0^t  \, V\st{t}(t')^2 + \underbrace{V\st{t}(t')^2 \cos(2\omega\st{c} t)} 
 dt'.
\label{eq:Delta_phi_Fts2}
\end{equation}
If $V\st{t}(t)$ varies slowly, the under-braced term integrates to zero over each cantilever oscillation period. 
In this case, the overall frequency shift $\Delta f = d \Delta \phi / dt$ caused by the first term is 
\[
 \Delta f\st{const}(t) = -\frac{f\st{c}}{4 k\st{c}} C''_q V\st{t}(t)^2.
\]
For the second term, we take the same basic approach. The second term simplifies to
\begin{equation}
\Delta \phi\st{osc} = - \frac{f\st{c} C\st{tip}'}{k\st{c} A_0 C\st{tip}} \int_0^t  V\st{t} q\st{t}^{(1)} \cos(\omega\st{c} t') dt'
\label{eq:delta_phi_2nd}
\end{equation}
We need the first-order tip-charge $q\st{t}^{(1)}$. In the time-independent case, this was given by
\begin{equation}
q\st{t}^{(1)} = C\st{tip} \, h \ast V_x = C\st{tip} \int_{t_0}^t h(t-t') V_x(t') \, dt',
\label{eq:first-order-tip-charge}
\end{equation}
where $h$ is the impulse response function between the applied tip-sample voltage $V\st{ts}(t)$ and the voltage that drops across the tip capacitor $V\st{t}=q\st{t}^{(0)}/C\st{tip}$, $\ast$ denotes convolution, and $V_x$ is the oscillating voltage induced by the oscillating tip displacement, $V_x = C\st{tip}' q\st{t}^{(0)} x^{(0)} / C\st{tip}^2$.
In the time-dependent case, the convolution integral is replaced by the generalized convolution
\begin{align}
q\st{t}^{(1)}& = C\st{tip} \int_{t_0}^{t} h(t, t') V_x(t') \, d t'. \nonumber \\
\intertext{The time-varying impulse response function $h(t,t')$ describes the response at time $t$ to an impulse applied at a time $t'$. Substituting in for $V_x$, we obtain}
q\st{t}^{(1)}& = \int_{t_0}^{t} h(t, t') \frac{q\st{t}^{(0)}(t')}{C\st{tip}} C\st{tip}' \, x^{(0)}(t')  \, d t'. \nonumber \\
\intertext{Substituting in for $x^{(0)}$ and simplifying gives}
q\st{t}^{(1)}& = A_0 C\st{tip}' \int_{t_0}^{t} h(t, t') V\st{t}(t') \cos(\omega\st{c} t')  \, d t'. 
\label{eq:first-order-tip-charge-time}
\end{align}
In the experiments we consider, the spectral content of $V\st{t}$ will be concentrated at low frequencies. Looking back to Eq.~\ref{eq:delta_phi_2nd}, the integral will only have non-zero values over a cantilever period when the response is in phase with the oscillating position. Therefore, to a good approximation, we need the component of the oscillating tip charge at the cantilever frequency $\omega\st{c}$. Since the spectral content of $V\st{t}$ is concentrated at low frequencies, this will be given by the \emph{time-varying} response function $H(\omega\st{c}, t)$
\begin{equation}
q\st{t}^{(1)} \approx A_0 C\st{tip}' V\st{t}(t) \left [ \Re{\! \big[ H(\omega\st{c}, t)\big]} \cos(\omega\st{c} t) - \Im{\! \big[ H(\omega\st{c}, t) \big] \sin(\omega\st{c} t)}  \right ]
\end{equation}
Substituting this equation for $q\st{t}^{(1)}$ into Eq.~\ref{eq:delta_phi_2nd}, we obtain
\[
\Delta \phi\st{osc} = - \frac{f\st{c} (C\st{tip}')^2}{k\st{c} C\st{tip}} \int_0^t V\st{t}(t')^2 \left [  \cos^2(\omega\st{c} t') \Re{\! \big[H(\omega\st{c}, t'
) \big ]} +  \sin(\omega\st{c} t') \cos(\omega\st{c} t') \Im{\! \big[H(\omega\st{c}, t') \big ]}  \right ] dt'
\]
Simplifying using the double angle formulas and the definition $\Delta C'' = 2 (C\st{tip}')^2/C\st{tip}$,
\begin{multline}
\Delta \phi\st{osc} = - \frac{f\st{c} \Delta C''}{4 k\st{c}}  \int_0^t  V\st{t}(t')^2  \bigg [ \underbrace{ \Re{\! \big[H(\omega\st{c}, t')\big]}} +  \cos(2\omega\st{c} t') \Re{\! \big[H(\omega\st{c}, t')\big]}  \\
+ \sin(2 \omega\st{c} t') \Im{\! \big[ H(\omega\st{c}, t') \big]}  \bigg] dt'.
\end{multline}
If $V\st{t}^2$ varies slowly and $H(\omega\st{c}, t)$ is linear within each cantilever period, then only the underbraced term contributes to the integral over a cantilever oscillation period.
In this case, the frequency shift $\Delta f = d \Delta \phi / d t$ is
\[
\Delta f\st{osc}(t) = -\frac{f\st{c} \Delta C''}{4 k\st{c}} V\st{t}(t)^2 \Re{\! \big[H(\omega\st{c}, t)\big]}.
\]
Adding the two terms, we obtain the overall frequency shift given in Eq.~\ref{eq:frequency_shift_time_varying}:
\begin{equation*}
\Delta f(t) = -\frac{f\st{c} }{4 k\st{c}} \left ( C''_q + \Delta C'' \Re{\! \big[H(\omega\st{c}, t)\big]}\right ) V\st{t}(t)^2.
\end{equation*}
The zeroth order tip voltage $V\st{t}(t)$ is 
\begin{equation}
V\st{t}(t) = \int_{t_0}^{t} h(t, t') V\st{ts}(t') \, d t' .
\label{eq:Vt_t_h}
\end{equation}
The two key approximations made in this derivation were (1) $V\st{t}$ and $V\st{t}^2$ do not have significant spectral content at the cantilever resonance frequency $\omega\st{c}$ and (2) the real and imaginary components of $H(\omega\st{c}, t)$ are linear over each cantilever period.

\section{Determining the time-varying response function}
The time-varying response function is the Fourier transform of the modified time-varying impulse response $\bar{h}(t,\tau)$, where $\tau = t-t'$ is the delay between the time $t$ at which the response is measured and the time $t'$ at which the impulse was applied. The modified time-varying impulse response is
\begin{align}
    \bar{h}(t, \tau)& = h(t, t-\tau)
\end{align}
and the Fourier transform is
\[
H(\omega, t) = \int_{-\infty}^{\infty} \bar{h}(t, \tau) e^{-j\omega \tau} \, d\tau
\]
Next we describe how $H(\omega\st{c}, t)$ can be calculated for the single parallel sample resistance and capacitance model used in the text ($R\st{s} \parallelslant C\st{s}$, circuit shown in Fig.~\ref{fig:df_t_explained}a).


In this case, the differential equations describing the evolution of the tip charge $q\st{t}$ can be expressed in terms of $q_R$, where $\dot{q}_R$ is the current through the sample resistance $R\st{s}$. The state variable $q_R$ is described by the differential equation
\begin{align}
\dot{q}_R& = -\omega\st{fast}(t) q_R + C\st{tip} \omega\st{fast}(t) V\st{ts}(t) 
\end{align}
where $\omega\st{fast}(t) = \big(R\st{s}(t) (C\st{s} + C\st{tip}) \big)^{-1}$.
To zeroth order, the dependence of the tip capacitance on distance is negligible and the system is linear. The system is time-varying through the time-dependence of the resistance $R\st{s}$. The tip charge and tip voltage are
\begin{align}
q\st{t}& = \left ( \frac{C\st{tip}}{C\st{tip} + C\st{s}}\right ) q_R + \left( \frac{C\st{tip} C\st{s}}{C\st{tip} + C\st{s}}\right )V\st{ts}(t) 
\label{eq:qt}
\\
V\st{t}& = \left ( \frac{1}{C\st{tip} + C\st{s}} \right ) q_R + \left( \frac{C\st{s}}{C\st{tip} + C\st{s}}\right )V\st{ts}(t)
\label{eq:Vt}
\end{align}
The propagator (or state transition matrix) $\Phi$ describes the evolution of the state variable, with $q_R(t) = \Phi(t, t_0) q_R(t_0)$ in the absence of a voltage input. For this system, the propagator is
\begin{equation}
\Phi(t,t') = e^{-\int_{t'}^{t} \omega\st{fast}(\theta) d\theta}.
\label{eq:Phi}
\end{equation}
The time-varying impulse response\footnote{From Kailath Ch. 9, Eq. 23 \cite{Kailath1980book}, $h(t, t') = C(t) \Phi(t, t') B(t') + D(t)\delta(t-t')$, where $B$, $C$, $D$ have their usual definitions for state space representations.} gives the response of the tip voltage at a time $t$ to a voltage input applied at time $t'$
\[
h(t,t') = \left ( \frac{C\st{tip}}{C\st{s} + C\st{tip}} \right) \omega\st{fast}(t') \Phi(t, t') u(t-t') + \left ( \frac{C\st{s}}{C\st{s} + C\st{tip}} \right ) \delta(t-t'),
\]
where $u(t)$ is the Heaviside step function ($u(t) = 0$ for $t<0$, $u(t) = 1$ for $t>0$) and $\delta(t)$ is the Dirac delta function.
The modified time varying impulse response is given by defining the delay $\tau = t-t'$,
\[
\bar{h}(t,\tau) = \left ( \frac{C\st{tip}}{C\st{s} + C\st{tip}} \right) \omega\st{fast}(t-\tau) \Phi(t, t-\tau) u(\tau) + \left ( \frac{C\st{s}}{C\st{s} + C\st{tip}} \right ) \delta(\tau).
\]
The sought-after time-varying frequency response is the Fourier transform of $\bar{h}$ with respect to $\tau$\footnote{From Shmaliy Eq. 6.31 \cite{Shmaliy2007book}, $H(\omega, t) = \bar{H}(\omega, t)$.}
\begin{align*}
H(\omega, t)& = \int_{-\infty}^{\infty} \left [ \left ( \frac{C\st{tip}}{C\st{s} + C\st{tip}} \right) \omega\st{fast}(t-\tau) \Phi(t, t-\tau) u(\tau)  + \left ( \frac{C\st{s}}{C\st{s} + C\st{tip}} \right ) \delta(\tau) \right ]e^{-j \omega \tau} \, d\tau, \\
& = \int_{0}^{\infty} \left ( \frac{C\st{tip}}{C\st{s} + C\st{tip}} \right) \omega\st{fast}(t-\tau) \Phi(t, t-\tau) e^{-j \omega \tau} \, d\tau  + \int_{-\infty}^{\infty} \left ( \frac{C\st{s}}{C\st{s} + C\st{tip}} \right ) \delta(\tau) e^{-j \omega \tau} d \tau.
\label{eq:time-varying-response1}
\end{align*}
Simplifying the second integral and substituting the expression for $\Phi$ from Eq.~\ref{eq:Phi}, we obtain Eq.~\ref{eq:time-varying-response}.
In words, $H(\omega, t)$ gives the response of $V\st{t}$ to an applied external voltage $e^{j \omega t}$. In the limit that $\omega\st{fast}$ is constant, $H(\omega, t)$ becomes time-independent and reduces to $H(\omega)$ given by Eq.~\ref{eq:H-full}.

\section{tr-EFM and pk-EFM Simulations}
\label{trEFMandpkEFM}
The approximation for $\Delta f$ given by Eq.~\ref{eq:frequency_shift_time_varying} was compared against the results of numerical simulations of the cantilever's dynamics for tr-EFM and pk-EFM experiments. 
To simulate a tr-EFM or pk-EFM experiment in which the light is turned on at $t=0$, the sample resistance was taken to respond to light with a time constant $\tau\st{L}$,
\begin{equation}
R\st{s}(t) = 
\begin{cases}
R\st{dark} & t\leq 0 \\
R\st{dark} + (R\st{light} - R\st{dark})(1-e^{-t/\tau\st{L}})& t> 0,
\end{cases}
\label{eq:Rs_t}
\end{equation}
where $R\st{dark}$ is the sample's dark resistance and $R\st{light}$ is the sample's final resistance after the light has been on for a long time.

To simulate a tr-EFM experiment, the tip-sample voltage was left constant $V\st{ts}(t) = V$. To simulate a pk-EFM experiment, the tip-sample voltage $V\st{ts}$ was stepped back to zero after a time $t\st{p}$ 
\begin{equation}
V\st{ts}(t) = \begin{cases}
V & t \leq t\st{p} \\
0 & t > t\st{p}.
\end{cases}
\label{eq:Vts}
\end{equation}
In our simulations, the tip-sample voltage was always $V = \SI{10}{\V}$. 

\begin{figure}
\includegraphics{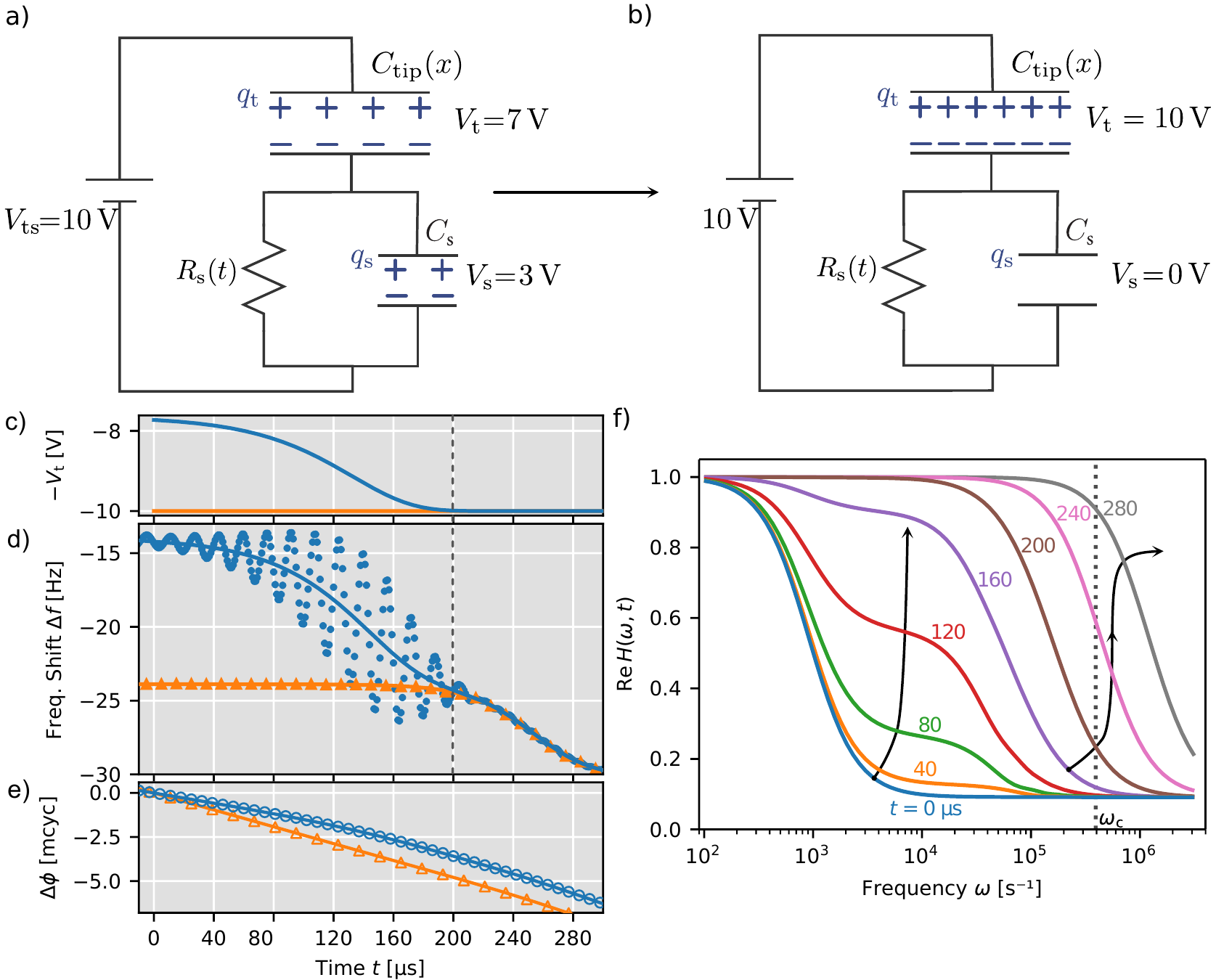}
\caption[Frequency shift and time-varying transfer function versus time for a representative pk-EFM experiment]{Frequency shift and time-varying transfer function versus time for a representative pk-EFM experiment. 
(a) The model circuit used in this simulation. After the tip-sample voltage is applied, charge builds up on the tip and sample capacitors.
(b) Eventually, the sample capacitor discharges and $V\st{t} = V\st{ts}$.
 The (c) negative zeroth order tip voltage $-V\st{t}$, (d) frequency shift, and (e) phase shift calculated using numerical simulations (points) and analytically using Eq.~\ref{eq:frequency_shift_time_varying} (line).
 For (c)--(e), $V\st{t}(t_0) = \SI{7}{\V}$ (blue circles), \SI{10}{\V} (orange triangles).
 (f) The time-varying response function $H(\omega,t)$ shown at times $t=$ 0, 40, 80, \ldots, \SI{280}{\us}. 
 Constant experimental parameters: $R\st{dark} = \SI{10}{\tera\ohm}$, $R\st{light} = \SI{10}{\mega\ohm}$, $\tau\st{L} = \SI{40}{\us}$, $k\st{c} = \SI{3.5}{\N\per\m}$, $f\st{c} = \SI{62}{\kilo\Hz}$, $Q=\num{26000}$, $C\st{tip} = \SI{1e-4}{\pF}$, $C\st{tip}' = \SI{-2.8e-5}{\pF\per\um}$, $C\st{s} = \SI{1e-5}{\pF}$, $C\st{tip}'' = \SI{6.77e-5}{\pF\per\um\squared}$, $V\st{ts} = \SI{10}{\V}$, $t_0 = \SI{-200}{\us}$.}
\label{fig:df_t_explained}
\end{figure}

In Eq.~\ref{eq:frequency_shift_time_varying}, the two factors that affect the frequency shift are the zeroth order tip voltage $V\st{t}$ and the time-varying transfer function $H(\omega\st{c}, t)$ at the cantilever frequency $\omega\st{c}$. Figure~\ref{fig:df_t_explained} illustrates the effect of each factor for a sample where the light decreases the sample resistance from $R\st{dark} = \SI{100}{\giga\ohm}$ to $R\st{light} = \SI{10}{\mega\ohm}$ with an exponential risetime of $\tau\st{L} = \SI{10}{\us}$. These values were chosen to clearly illustrate how each factor affects the frequency shift.

The experiment begins when the applied voltage $V\st{ts}$ is switched from 0 to \SI{10}{\V}. Initially, equal charges build up on both capacitors and the voltage drop across the tip capacitor is $V\st{t} = V\st{ts} C\st{s}/(C\st{s} + C\st{tip})$ (Fig.~\ref{fig:df_t_explained}a). The sample capacitor discharges with an $RC$ time constant $R\st{dark}(C\st{tip} + C\st{s}) = \SI{1.1}{\ms}$, so that eventually, all of the applied voltage drops across the tip capacitor (Fig.~\ref{fig:df_t_explained}b).
When the light is turned on, the sample resistance $R\st{s}$ decreases by 6 orders of magnitude, decreasing the $RC$ time constant from \SI{1.1}{\ms} to \SI{1.1}{\ns}.

Dramatic differences in the frequency shift are observed depending on when the light is turned on.
Fig.~\ref{fig:df_t_explained}c shows the tip voltage versus time after the light is turned on ($t=0$) when the tip is only partially charged (blue), and, for comparison, when the tip is fully charged (orange).
If the tip is only partially charged, the light-induced decrease in sample resistance speeds up the charging of the tip capacitor---the tip charges fully in \SI{200}{\us} compared to the \SI{3}{\ms} it would take in the dark.
If allowed to charge fully before turning on the light, $V\st{t}$ remains constant (orange).

Figure~\ref{fig:df_t_explained}d shows the resulting frequency shift versus time signal calculated from numerical simulations (points) and approximated using Eq.~\ref{eq:frequency_shift_time_varying} (lines).
According to Eq.~\ref{eq:frequency_shift_time_varying}, the frequency shift $\Delta f$ is proportional to $V\st{t}^2$, so the blue trace and circles show an increase in frequency shift from $t=0$ to \SI{200}{\us} as the tip charges.
The messy oscillations in the numerically simulated frequency-shift transient (blue circles) occur because the tip charge and dc displacement are changing rapidly on the timescale of the cantilever period, so the cantilever frequency shift is poorly defined during these periods.
The phase shift is still well-defined, however, and Fig.~\ref{fig:df_t_explained}d shows that Eq.~\ref{eq:frequency_shift_time_varying} accurately captures the numerically simulated phase shift remarkably well.

The increase in frequency shift after \SI{200}{us} is caused by an increase in the time-varying transfer function at the cantilever frequency ($H(\omega\st{c}, t)$). Physically, as the circuit's $RC$ time constant drops below the cantilever's inverse resonance frequency $\omega\st{c}^-1 = \SI{2.5}{\us}$, more of the tip charge is able to oscillate on and off the tip during motion, so that the oscillating force in-phase with the tip displacement increases. 
The predicted frequency and phase shift agrees closely with the numerical simulations.

All of these light and bias-dependent effects can be understood using the time-varying transfer function $H(\omega, t)$.
Fig.~\ref{fig:df_t_explained}f plots the real part of $H$ versus angular frequency at \SI{40}{\us} intervals after the light is turned on at $t=0$. 
Over the first \SI{160}{\us} after the light is turned on, the main effect of the reduced sample resistance is an increase in $H$ at frequencies well below the cantilever frequency $\omega\st{c}$.
This low-frequency response can affect the frequency shift through the tip voltage $V\st{t}$, which is the generalized convolution of the time-varying impulse response function and the applied tip-sample voltage (Eq.~\ref{eq:Vt_t_h}).
Notice that the increase in $H$ at low-frequencies has no effect if all of the tip-sample voltage is already dropping across the tip, as for the orange traces in Fig.~\ref{fig:df_t_explained}(c,d).
In this case, light-induced frequency shifts can be attributed to changes in $\Re{H(\omega\st{c}, t)}$ alone.
The experimental data of Fig.~\ref{fig:trEFM-tp-time} shows this same dependence of the light-induced frequency shift on the initial tip charge.

While the effect of the increase in $H$ at low-frequencies depends on the initial tip charge, a change in $\Re{H(\omega\st{c}, t)}$ directly affects the cantilever frequency shift. Between \SI{160}{\us} and \SI{280}{\us}, the real part of the transfer function $\Re{H(\omega\st{c}, t)}$ increases dramatically at the cantilever frequency (Fig.~\ref{fig:df_t_explained}a).
The cantilever frequency shift (Fig.~\ref{fig:df_t_explained}c) increases further as a result.

\begin{figure}
\includegraphics{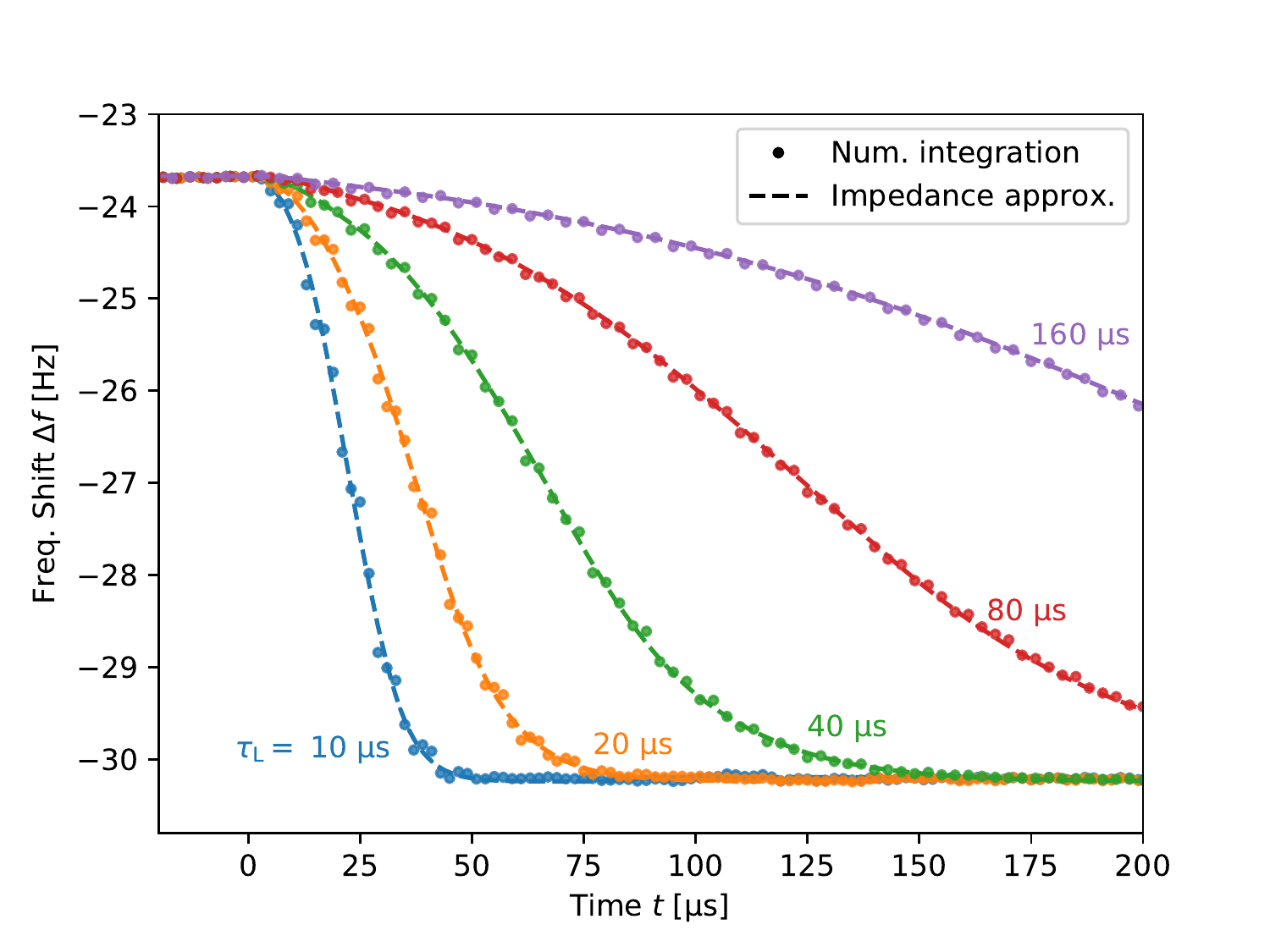}
\caption[Frequency shift versus time for representative simulated EFM experiments.]{Frequency shift versus time for representative simulated EFM experiments. For different light-induced time constants $\tau\st{L}$ (Eq.~\ref{eq:Rs_t}), points show the frequency shift calculated from numerical integration of Eq.~\ref{eq:dot_x_num_int} and dashed lines show the results calculated using the approximation in Eq.~\ref{eq:frequency_shift_time_varying}. For each trace, the following experimental parameters were held constant: $f\st{c} = \SI{62.5}{\kHz}$, $k\st{c} = \SI{3.5}{\N\per\m}$, $Q = \num{26000}$, $A_0 = \SI{50}{\nm}$, $\phi_0 = \pi$, $t_0 = \SI{-100}{\us}$, $R\st{dark} = \SI{100}{\giga\ohm}$, $R\st{light} = \SI{10}{\mega\ohm}$, $C\st{s} = 0$, $C\st{tip} = \SI{1e-4}{\pF}$, $C''_q = \SI{52.1}{\pF\per\nm\squared}$, $\Delta C'' = \SI{15.7}{\pF\per\nm\squared}$, $V\st{t}(t_0) = \SI{10}{\V}$, $V\st{ts}= \SI{10}{\V}$.}
\label{fig:simulated-EFM}
\end{figure}

Simulations were performed for a wide variety of experimental parameters, designed to cover regimes where each factor ($V\st{t}$ and $\Re{H(\omega\st{c}, t)}$) influences the frequency shift. For sample tr-EFM experiments, Figure~\ref{fig:simulated-EFM} shows that Eq.~\ref{eq:frequency_shift_time_varying} is a good approximation of the cantilever frequency shift.

\begin{figure}
\includegraphics[width=6in]{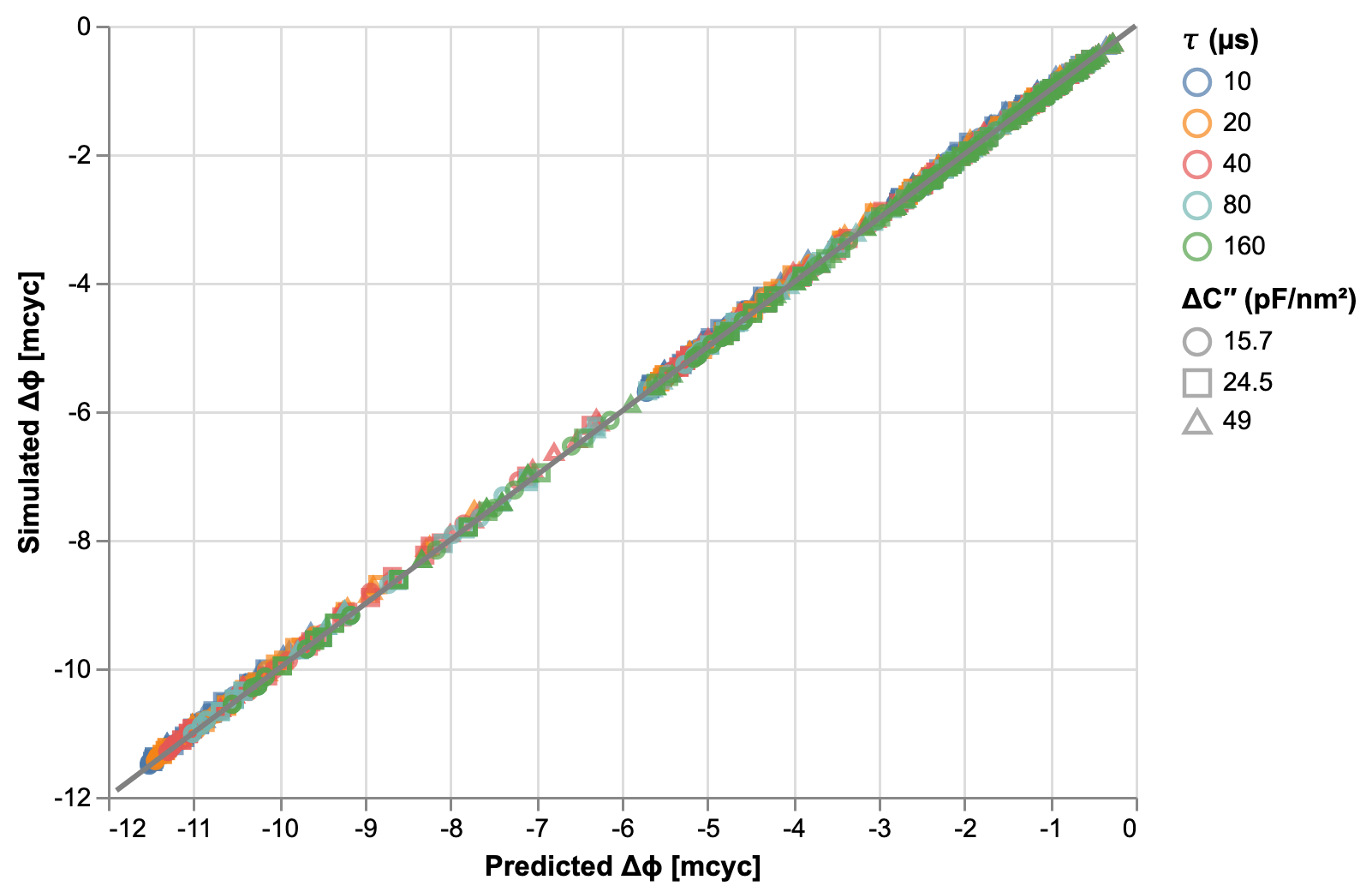}
\caption[Simulated phase shift agrees closely with predicted phase shift in pk-EFM experiments.]{Simulated phase shift agrees closely with predicted phase shift in pk-EFM experiments. Experimental parameters:
$R\st{dark}$ = \num{1e5}, \num{1e7}, \SI{1e8}{\mega\ohm};
$R\st{light} = \SI{10}{\mega\ohm}$; 
$\tau\st{L} =$ 10, 20, 40, 80, \SI{160}{\us}; 
$V\st{t}(t_0) =$ 7, \SI{10}{\V};
$t\st{p} = $ 48, 96, 192, \SI{384}{\us};
$C\st{s} = $ \num{0.01}, \num{0.04}, \SI{0.1}{\fF};
$C\st{tip}' = $ \num{-28}, \num{-35}, \SI{-49.5}{\fF\per\nm};
$C\st{tip} = $ \SI{0.1}{\fF};
$f\st{c} = \SI{62.5}{\kHz}$;
$k\st{c} = \SI{3.5}{\N\per\m}$;
$Q = \num{26000}$, $A_0 = \SI{50}{\nm}$;
$A_0 = \SI{50}{\nm}$;
$t_0 = \SI{-200}{\us}$;
$\phi_0 = 0$.
}
\label{fig:compare-phase-shift}
\end{figure}

To assess pk-EFM experiments, the measured and predicted phase shifts were compared for 1080 simulated experiments. Figure~\ref{fig:compare-phase-shift} shows that over a wide variety of sample and cantilever parameters, the predicted and measured phase shift in pk-EFM experiments agreed closely. The residuals $r = \Delta \phi\st{simulated} - \Delta \phi\st{predicted}$ had a mean and standard deviation of $\SI{0.027}{\milli\cycle}$ and $\SI{0.049}{\milli\cycle}$ respectively. The maximum residual was $\SI{0.20}{\milli\cycle}$.

\subsection{Numerical simulations}
\label{sec:numerical_simulations}
The numerical simulations used to generate Figs.~\ref{fig:simulated-EFM}--~\ref{fig:compare-phase-shift} were performed in Python. The code for these simulations is publicly available \cite{Dwyer2020may_data}. 
The coupled first-order ordinary differential equations that described the cantilever dynamics are
\begin{subequations}\label{eq:dot_x_num_int}
\begin{align}
\dot{x}& = v \\
\dot{v}& = -\omega\st{c}^2 x - \frac{\omega\st{c}}{Q} v + \frac{C\st{tip}'(x)}{2mC\st{tip}(x)^2} q\st{t}^2 
\label{eq:v}
\\
\dot{q}_R& = -\frac{1}{R\st{s}(t)(C\st{tip}(x) + C\st{s})}q_R + \frac{C\st{tip}(x)}{R\st{s}(t)(C\st{tip}(x) + C\st{s})} V\st{ts}(t).
\end{align}
\end{subequations}
where $v$ is the cantilever velocity.
For convenient comparison to the model developed above, the zeroth order charge was also computed using
\begin{equation}
\dot{q}_{R}^{(0)} = -\frac{1}{R\st{s}(t)(C + C\st{s})}q_{R}^{(0)} + \frac{C}{R\st{s}(t)(C + C\st{s})} V\st{ts}(t),
\end{equation}
where $C=C\st{tip}(0)$.
The tip capacitance and its derivatives were given by $C\st{tip}(x) = C + C' x + C'' x^2/2$.
In Eq.~\ref{eq:v}, the tip charge $q\st{t}$ was calculated using Eq.~\ref{eq:qt}.
The numerical integration was performed in Python using Scipy's odeint function, storing the state vector $\vc{y} = (x\,\,\, v\,\,\, q_R\,\,\, q_R^{(0)})^T$ every \SI{1}{\us}.

The initial state vector for the cantilever was
\begin{equation*}
\vc{y}_0 = \begin{pmatrix}
x\\ 
v\\ 
q_R \\
q_R^{(0)}
\end{pmatrix}
=
\begin{pmatrix}
A_0 \cos( \omega\st{c} t_0 + \phi_0) \\
A_0 \omega\st{c} \sin( \omega\st{c} t_0 + \phi_0) \\
\beta \, C V \\
\beta \, C V
\end{pmatrix},
\end{equation*}
where $A_0$ is the cantilever's initial amplitude, $\phi_0$ is the cantilever's initial phase, $t_0 = \SI{-100}{\us}$ is the initial time the numerical integration was started, $\beta$ is the initial fraction of the tip sample voltage that drops across the tip capacitor, and the applied tip-sample voltage $V = \SI{10}{\V}$ (Eq.~\ref{eq:Vts}).

 The output of the numerical integration was used to determine the cantilever's amplitude, phase, and frequency. The cantilever amplitude $A = \abs{z}$ and phase $\phi = \arg{z}$ were calculated using the complex number
\begin{equation}
z = (x - x\st{eq}) - j v/\omega\st{c}. 
\end{equation}
where $x\st{eq} = C\st{tip}' V\st{ts}^2/(2 k\st{c})$.
The amplitude and phase data were filtered by averaging over a single cantilever period (16 data points).
The frequency shift was calculated by numerically differentiating the filtered phase shift (using second order central differences via numpy's gradient function).

To compare the impedance theory approximation to the numerical integration, the zeroth order tip voltage $V\st{t}$ and the time-varying frequency response $H(\omega, t)$ must be known.
The zeroth order tip voltage was calculated using Eq.~\ref{eq:Vt} with $q_R = q_R^{(0)}$.
To approximate $H(\omega, t)$, the double integral of Eq.~\ref{eq:time-varying-response} was broken into pieces depending on the value of $t$ and $\omega\st{fast}(t)$.
When $\omega\st{fast} \ll \omega\st{c}$, the necessary integral is oscillatory and decays slowly which makes converging a numerical approximation difficult. 
In this case, the integrand was sampled at points equally spaced through the cantilever cycle (\num{160} points per cycle) and Simpson's rule was used to compute the value of the integral.

\section{Scanning probe microscopy}
The scanning probe microscopy set up used to perform different measurements here has been described in our previous reports \cite{Dwyer2017jun}.
Cantilever motion was detected using a fiber interferometer operating at \SI{1490}{\nano\meter} (Corning SMF-28 fiber).
The laser diode’s (QPhotonics laser diode QFLD1490-1490-5S) dc current was set using a precision current source (ILX Lightwave LDX-3620), and the current was modulated at radio frequencies using the input on the laser diode mount (ILX Lightwave LDM 4984, temperature-controlled with ILX Lightwave LDT-5910B).
The interferometer light was detected with a 200-kHz bandwidth photodetector (New Focus model 2011, built-in high-pass filter set to \SI{200}{\kilo\hertz}) and digitized at \SI{1}{\mega\hertz} (National Instruments, PCI-6259). 
The cantilever was driven using a commercial PLL cantilever controller (RHK Technology, PLLPro2 Universal AFM controller) with PLL feedback loop integral gain $I$ $=$ \SI{2.5}{\hertz}, proportional gain $P$ $=$ \SI{-5}{\degree\hertz}.
The sample was illuminated from above with a fiber-coupled $\SI{405}{\nano\meter}$ laser (Thorlabs model LP405-SF10, held at $\SI{25}{\celsius}$ with a Thorlabs model TED200C temperature controller).
The laser current was controlled using the external modulation input of the laser's current controller (Thorlabs model LDC202, $\SI{200}{\kilo\Hz}$ bandwidth).
The light was coupled to the sample through a multimode, $\SI{50}{\micro\meter}$ diameter core, $0.22$ NA optical fiber (Thorlabs model FG050LGA).
The intensity at sample surface was calculated based on an estimated spot size of $\approx$ \SI{0.26}{\milli\meter\squared}

Implementation of broad band local dielectric spectroscopy has been described previously in Refs.~\citenum{Tirmzi2017jan} and \citenum{Tirmzi2019jan}.
The procedure is reproduced below for reference.
For amplitude modulation BLDS (Fig.~\ref{fig:frequency-BLDS}b), we applied a time-dependent voltage to the cantilever tip :
\begin{equation}
V\st{m}(t) 
	= V\st{pp} 
	\left(  \frac{1}{2} + \frac{1}{2} \cos(2 \pi f\st{am} t) \right)
	\: \cos(2\pi f\st{m} t).
\label{eq:tip-modulation-BLDS}
\end{equation}
In the experiments reported in the manuscript, $f\st{m} = \SI{200}{\hertz}$ to $\SI{1.5}{\mega\hertz}$, $f\st{am} = \SI{45}{\hertz}$, and the amplitude was set to $V\st{pp} = \SI{6}{\V}$.
The time-dependent voltage in Equation~\ref{eq:tip-modulation-BLDS} was generated using a digital signal generator (Keysight 33600).
The cantilever frequency shift was measured in real time using a phase-locked loop (PLL; RHK Technology, model PLLPro2 Universal AFM controller), the output of which was fed into a lock-in amplifier (LIA; Stanford Research Systems, model $830$).
The LIA time constant and filter bandwidth were $\SI{300}{\ms}$ and $\SI{6}{\dB}/\mathrm{oct}$, respectively. 
At each stepped value of $f\st{m}$, a wait time of $\SI{1500}{\ms}$ was employed, after which frequency-shift data were recorded for an integer number of frequency cycles corresponding to $\approx \SI{2}{\sec}$ of data acquisition at each $f\st{m}$.
The measurable $\Delta f\st{BLDS}$ primarily probes the response at $\omega\st{m}$ (Eq.~\ref{eq:Deltaf-BLDS}).
\begin{equation}
\Delta f\st{BLDS}(\omega\st{m}) =
-\frac{f\st{c} V\st{m}^2}{16 k} \Big[
C''_q + \Delta C'' \Re \Big( \hat{H}(\omega\st{m} + \omega\st{c}) \\
 + \hat{H}(\omega\st{m} - \omega\st{c}) \big) 
\Big]
\abs{\hat{H}(\omega\st{m})}^2
	\label{eq:Deltaf-BLDS}
\end{equation}
$\Delta f\st{BLDS}$ is related to the plotted voltage-normalized frequency shift $\alpha$ by Eq.~\ref{eq:alpha-kfc}.
\begin{equation}
\alpha = \frac{\Delta f\st{BLDS}(\omega\st{m})}{V\st{m}^{2}}.
	\label{eq:alpha-kfc}
\end{equation}
The $\Delta f\st{BLDS}$ frequency-shift signal was obtained from the LIA outputs as follows.
From the (real) in-phase and out-of-phase voltage signals $V_X$ and $V_Y$, respectively, a single (complex signal) in hertz was calculated using the formula
\begin{equation}
Z\st{Hz} = \left (V_x + j \, V_y  \right) \frac{S}{10} \times \sqrt{2} \times 20 \, \frac{\text{Hz}}{\text{V}}
\end{equation}
From $Z\st{Hz}$ we calculate $\alpha$
\begin{equation}
\alpha = \frac{4 \lvert Z\st{Hz}\rvert}{V\st{pp}^2}
 = \frac{8 \sqrt{2} \, S \sqrt{V_x^2 + V_y^2}}{V\st{pp}^2}.
\label{eq:alpha-calculation}
\end{equation}
In frequency shift BLDS (Fig.~\ref{fig:frequency-BLDS}a) described in Ref.~\citenum{Tirmzi2019jan}, the applied waveform is not amplitude modulated at \SI{45}{\hertz} and instead an equal period ON/OFF amplitude-modulating is applied.
The resultant frequency shift is calculated by software demodulation of the cantilever response by subtracting the average frequency shift during the ON period from the OFF period.
 
\subsection{tr-EFM and pk-EFM}
Implementation and data work up details for tr-EFM and pk-EFM have been extensively described in Ref.~\citenum{Dwyer2017jun} and Ref.~\citenum{Dwyer2017aug} and is reproduced briefly below for reference.
A commercial pulse and delay generator (Berkeley Nucleonics, BNC565) was used to generate tip voltage and light modulation pulses, as well as to turn off the cantilever drive voltage.
The PLLPro2 generated the cantilever drive voltage and also output a \SI{1}{\volt}, phase-shifted sine wave copy of the cantilever oscillation, generated by an internal lock-in amplifier coupled to the phase-locked loop.
A home built gated cantilever clock circuit converted this \SI{1}{\volt} phase-shifted sine wave to a square wave, which was used as a clock for timing tip voltage and light pulses.
A \SI{5}{\volt} digital signal output by the National Instruments PCI-6259 gates the clock, controlling the start of the experiment.
The BNC565 was used to trigger all signals relative to the cantilever clock.
Cantilever drive was switched off (\SIrange{2}{10}{\milli\second}) before the start of light pulse.
The raw cantilever oscillation data (digitized at 1 MHz) was saved along with counter timings (PCI-6259, 80-MHz counter) indicating the precise starting time of the light pulse (synchronized to the cantilever oscillation), allowing the start of the light pulse to be determined to within \SI{12.5}{\nano\second}.
Along with each pk-EFM phase shift data point, a control data point, identical except without turning on the light, was collected.

\begin{figure}[t]
\includegraphics[width=3.25in]{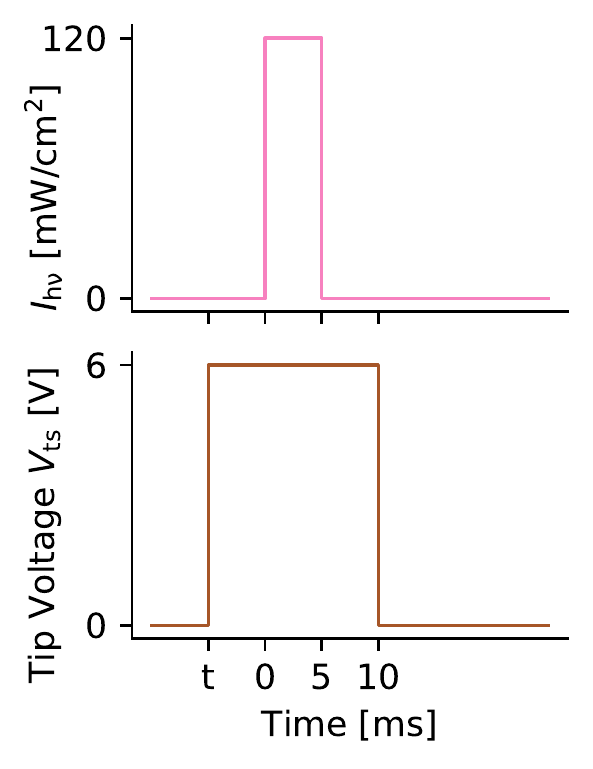}
\caption[Timing diagram for Fig.~\ref{fig:trEFM-tp-time}]{Timing diagram for voltage and light pulse used in Fig.~\ref{fig:trEFM-tp-time} experiments. Bias time in dark $t$ is varied from \SI{-2}{\milli\second} to \SI{-1000}{\milli\second}.}
\label{fig:tr-EFM-t}
\end{figure}

\begin{figure}[t]
\includegraphics[width=3.25in]{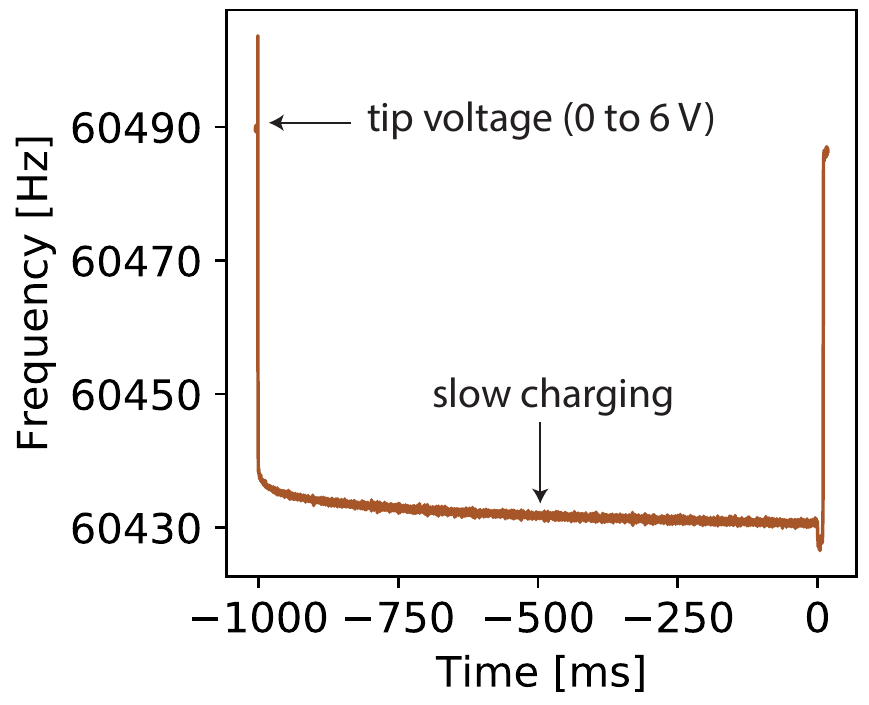}
\caption[tr-EFM with 1 sec $t\st{p}$ time]{tr-EFM frequency shift for $t\st{p}$ $=$ \SI{1}{\second} shown in Fig.~\ref{fig:trEFM-tp-time}. Additional frequency shift due to slow charging is apparent in the measured frequency shift after the tip voltage $V\st{ts}$ is changed to \SI{6}{\volt} even after several hundred millisecond wait.}
\label{fig:tr-EFM-1sec}
\end{figure}

\begin{figure}[t]
\includegraphics[width=3.25in]{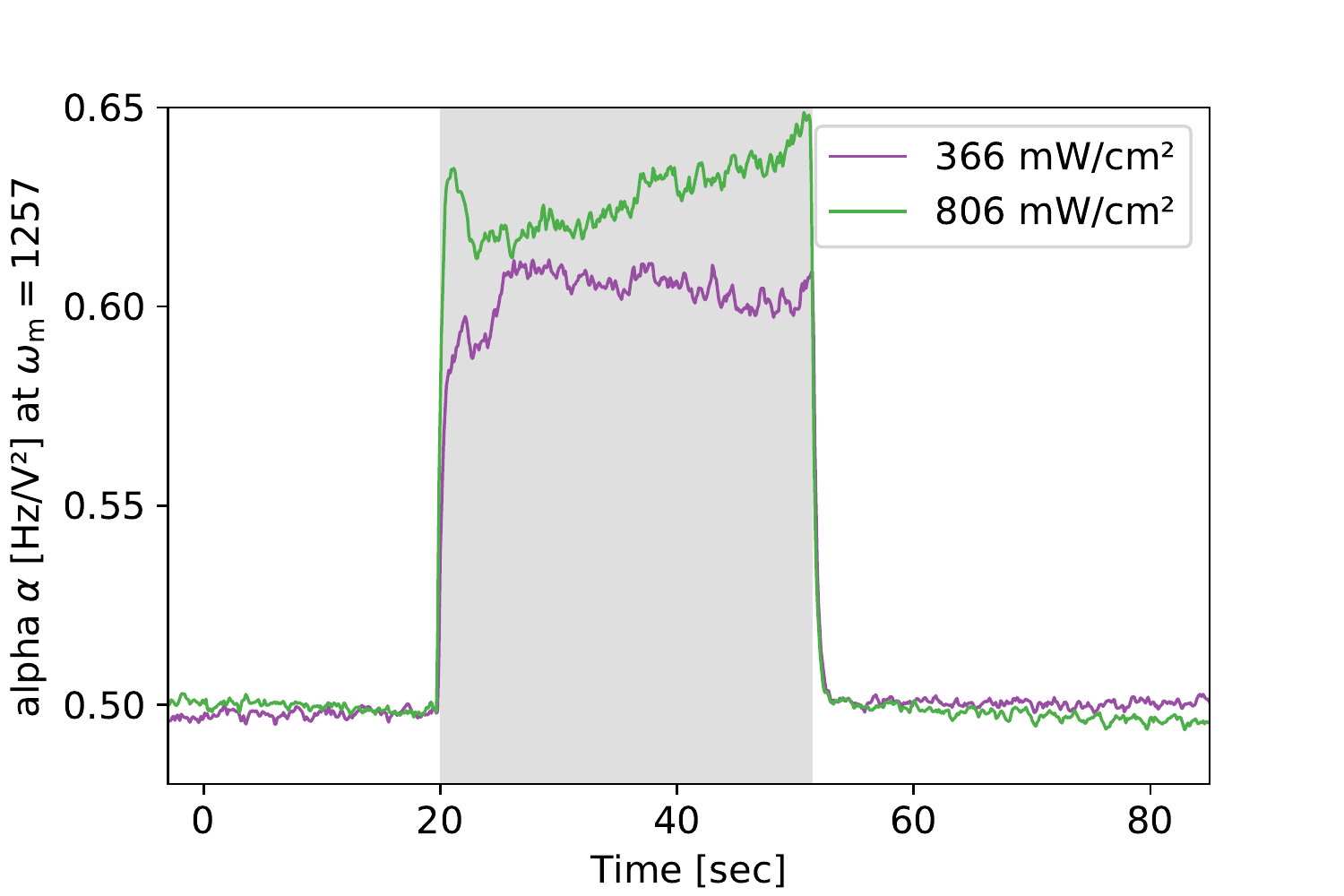}

\caption[Fixed frequency dielectric response at $\omega\st{m}$ $=$ \SI{1257}{\per\second}]{Dielectric response at $\omega\st{m}$ $=$ \SI{1257}{\per\second} for a period of illumination at two different light intensities. Dielectric response measured by $\alpha$ increases and decreases within the time resolution of the measurement ($\approx$ \SI{1}{\second}). Light was turned at time $t$ $=$ \SI{20}{\second} and turned off at time $t$ $=$ \SI{52}{\second}. Experimental parameters: $V\st{ts} = \SI{6}{\V}$, $h = \SI{200}{\nano\meter}$.}
\label{fig:fixed-freq-SI}
\end{figure}

\begin{figure}[t]
\includegraphics[width=3.25in]{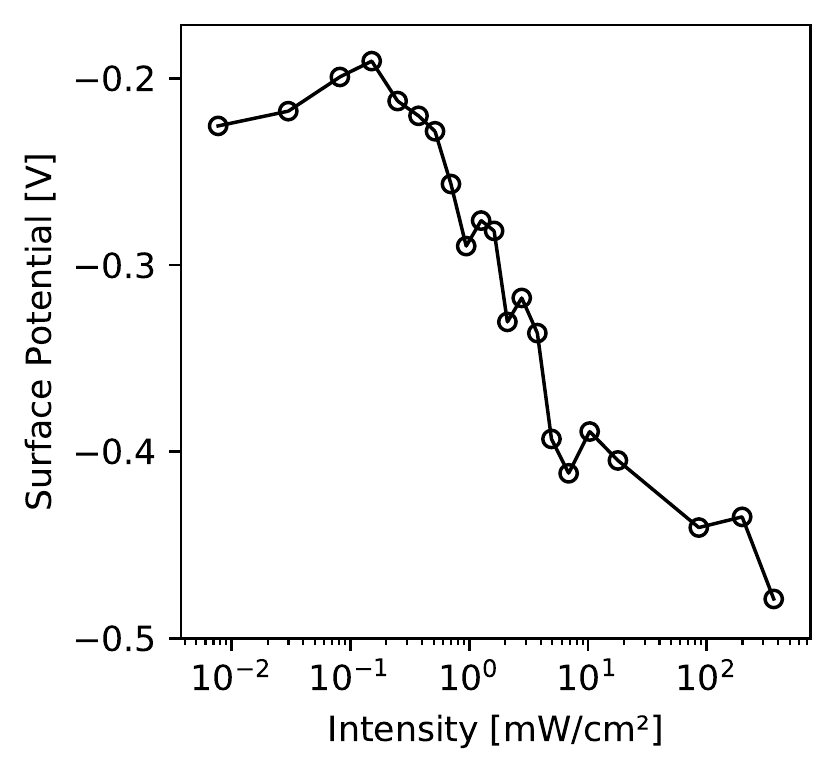}
\caption[Surface Potential]{Surface potential measured through frequency voltage parabolas at selected light intensities.}
\label{fig:surface-potential}
\end{figure}

\begin{figure}[t]
\includegraphics[width=3.25in]{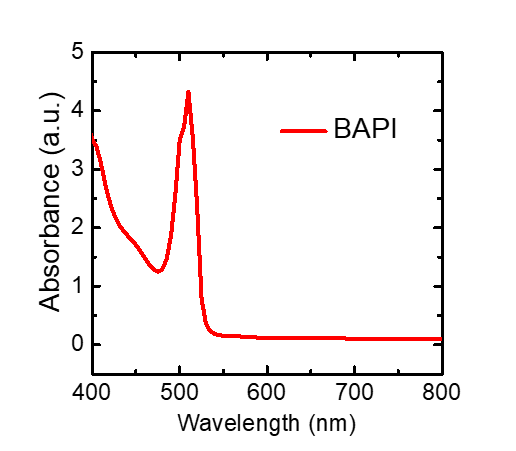}
\caption[UV-Vis]{Absorbance spectra of \ce{BA2PbI4}.}
\label{fig:absorbance}
\end{figure}
\begin{figure}[t]
\includegraphics[width=3.25in]{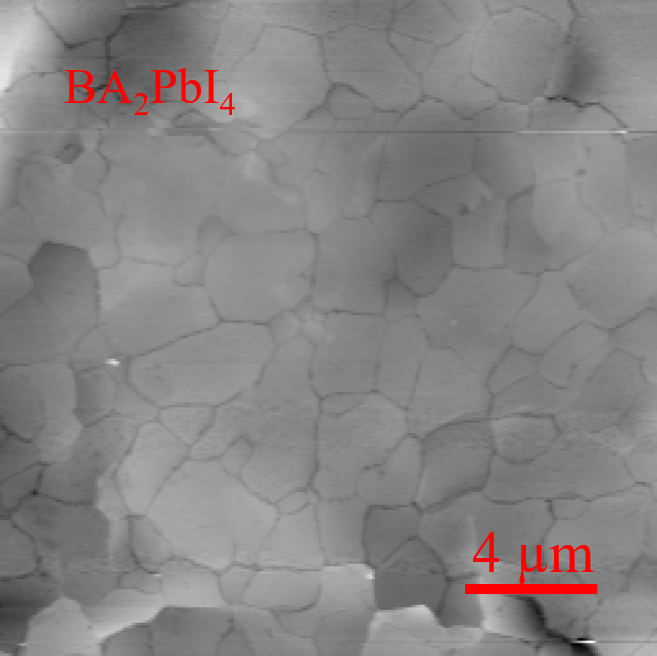}
\caption[AFM]{AFM topography image of the perovskite film.}
\label{fig:AFM}
\end{figure}

\begin{figure}[t]
\includegraphics[width=3.25in]{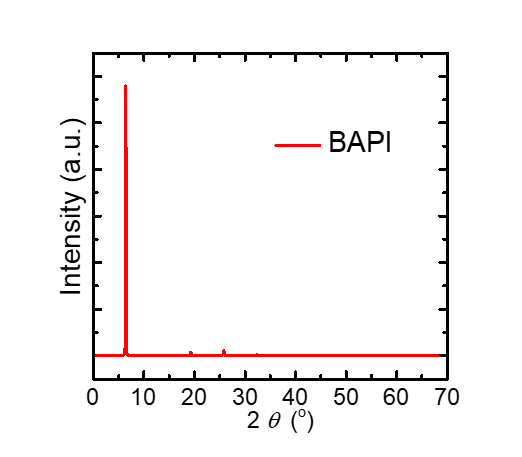}
\caption[XRD]{XRD spectra of the perovskite film.}
\label{fig:XRD}
\end{figure}



\clearpage
\listoffigures
\printfigures

\label{TheEnd-SI}